\documentclass[review,authoryear]{elsarticle}

\journal{Mathematical Psychology}

\biboptions{authoryear}

\newcommand\myFigureWidth{0.38}

\begin{document}

\begin{frontmatter}

\title{MODELING RESPONSE TIME WITH POWER-LAW DISTRIBUTIONS}



\author[mymainaddress]{Z. Liu}
\author[mysecondaryaddress]{O. Pavlov Garcia}
\author[mysecondaryaddress]{J. G. Holden}
\author[mymainaddress]{R. A. Serota\fnref{myfootnote}}
\fntext[myfootnote]{serota@ucmail.uc.edu}

\address[mymainaddress]{Department of Physics, University of Cincinnati, Cincinnati, Ohio 45221-0011}
\address[mysecondaryaddress]{Department of Psychology, CAP Center for Cognition, Action, and Perception,
University of Cincinnati, Cincinnati, Ohio 45221-0376}

%
%

\begin{abstract}
Understanding the properties of response time distributions is a long-standing
problem in cognitive science. We provide a tutorial overview of several
contemporary models that assume power law scaling is a plausible description of
the skewed heavy tails that are typically expressed in response time distributions.
We discuss several properties and markers of these distribution functions that have
implications for cognitive and neurophysiological organization supporting a given
cognitive activity. We illustrate how a power law assumption suggests that
collecting larger samples, and combining individual subjects' data into a single set
for a distribution-function analysis allows for a better comparison of a group of
interest to a control group. We demonstrate our techniques in contrasts of response
time measurements of children with and without dyslexia.
\end{abstract}


\end{frontmatter}


\section{Introduction}

Response time (RT) is the elapsed interval between the presentation of a stimulus and
the execution of a response in a laboratory-based cognitive task. Response times are
widely used in basic cognitive science and are also routinely used in
neuropsychiatric assessments for certain cognitive impairments and reading
disabilities \citep[e.g.][]{ancelin2006non, epstein2011effects, tse2010effects}. To date, there is little scientific consensus regarding the
fundamental statistical properties of response time distributions themselves. In
light of their widespread use, an accurate statistical description of response time
distributions would be extremely useful.

The focus of this article is response time probability density functions (PDFs) that
include skewed power-law tails. We tentatively assume that response time PDFs are
at least approximately stationary -- they stabilize near a steady state for a given
cognitive activity. This assumption is tentative because recent reports established
that time-series of response times express $1/f$ noise -- long-range correlations and
fractional dimensionality \citep{van2003self, van2005human}. Moreover, several
groups have proposed that the characteristic positive skew in tails of response time
distributions conforms to a power-law function \citep{holden2012self, ihlen2013influence}.

Previously, we examined the skewed tails of the probability distribution function of
response times \citep{holden2012self, holden2009dispersion, ma2015distribution}. 
Power-law scaling prevailed as a more plausible description of
empirical distribution's tail behavior than models that adopted exponentially
decaying tail behavior or even heavy-tailed lognormal behavior. In what follows, we
first introduce three candidate distributions that implement power-law tails: The
L$\acute{e}$vy alpha-stable distribution (S), the generalized inverse gamma distribution
(GIGa) and the "Cocktail" lognormal/Pareto (LNP) mixture distribution. Following
our introduction to the power-law distribution models, we use both the GIGa and
the LNP distributions to introduce heavy-tailed research design practices and other
methods for aggregating response time observations for the purpose of statistical
inference. Finally, we illustrate the use of these selected techniques on a published
data set that contrasted a sample of dyslexic and matched age-appropriate readers.

\section{Potential Model Distributions with Power-law Tails}

Here we summarize the properties of the potential model distributions with power-law 
tails. More detailed explanations of the role of the scale, shape, and location
parameters appear in Sec. \ref{Rescaling} in the context of the discussion of distribution
rescaling.

\subsection{The Stable Distribution}

With the exception of $\alpha = 2$, when it reduces to normal distribution (N), the S distribution \citep{nolan2015stable},
has a power-law tail
\begin{equation}
S(x; \alpha, \beta, \mu, \sigma) \sim |x|^{-(1 + \alpha)}
\end{equation}
Here stability $0 < \alpha < 2$ , defined below, and skewness $-ˆ'1 \le \beta \le 1$ are the shape
parameters of the distribution, while $\sigma$ and $\mu$ are its scale and location parameters.
In general, S does not have a closed-form expression -- among notable exceptions are
the Cauchy and L$\acute{e}$vy distributions -- instead, it is described by a closed-form
characteristic function and via parametric forms.

The two important properties of S are as follows. First, the sum of two
independent identical variables (iid) distributed with $S(x;∗,∗,∗, \sigma)$, is distributed
with $S(x;∗,∗,∗, 2^{1/\alpha} \sigma)$. That is, the linear combination of iid's results in the same
distribution, up to a rescaled scale parameter -- this is the definition of stability.
Second, the sum of independent identically distributed variables, whose PDF has an
$|x|^{-(1+\alpha)}$ tail, $0 < \alpha < 2$, tends to $S(x; \alpha, 0, 0, \sigma)$ as the number of variables in the
sum grows, by the Generalized Central Limit Theorem. These properties account for
the so called "Noah Effect" and resulting $H = 1/\alpha \ne 1/2$ Hurst exponent
\citep{mandelbrot2002gaussian}. Note, however, while $H = 1/2$ for $\alpha > 2$, according to CLT, the
number of variables in the sum and the corrections to N are abnormally large for fat
tails \citep{lam2011corrections}.

\citet{ihlen2013influence}, proposed S for fitting of RT distributions. While an attractive
proposition due to its properties above -- namely, power-law tails and additivity --
the two main drawbacks are that the power-tail exponents, $(1 + \alpha)$, are limited to
the $(1, 3)$ range and that the only allowed value of $\beta$ is $\beta = 1$, since otherwise the
variable of the distribution is not positively defined, 
\footnote{Notice similar limitations of the Fieller's and recinormal distributions \citep{moscoso2008theory}, 
whose power-law tail exponent is limited to a single value of $2$.}
as is physically meaningful for
RT. The allowed values of the exponent, however, contradict the observed exponent
values of several RT studies (see, for instance, \citep{holden2012self, ma2015distribution}).

\subsection{Generalized Inverse Gamma Distribution}

The application of GIGa to RT distribution fitting is motivated by its origins in the
generalized Bouchaud-M$\acute{e}$zard network model \citep{bouchaud2000wealth} -- and
its implications for brain dynamics -- and a related stochastic differential equation
describing a stochastic "birth-death" model \citep{ma2015distribution, ma2013distribution, ma2014model}. 
GIGa is zero for $x < \mu$, while for $x > \mu$ it is given by
\begin{equation}
GIGa(x; \alpha, \beta, \gamma, \mu) = \frac{\exp\left( 
-\left( \frac{\beta}{x-\mu}\right)^{\gamma} \right)
\gamma \left(\frac{\beta}{x-\mu} \right)^{\alpha \gamma + 1}}{\beta \Gamma(\alpha)}
\end{equation}
Above, $\alpha$ and $\gamma$ are the shape parameters, $\beta$ is the scale parameter, $\mu$ is the location
parameter and $\Gamma$ is the Gamma function. Except for $\mu$, all parameters are positively
defined and, in the context of RT, $\mu$ must be non-negative as well. For large $x$, it exhibits
power-law tail
\begin{equation}
\label{GIGaTailExponent}
GIGa(x; \alpha, \beta, \gamma, \mu) \sim x^{-(1 + \alpha \gamma)}
\end{equation}
Unlike S, there is no a priori upper bound on the tail exponent of GIGa.

\subsection{Lognormal/Pareto Mixture Distribution}

In the cocktail model \citep{holden2012self, holden2009dispersion}, 
the LNP mixture distribution is given by
\footnote{While using slightly different notations, this form is equivalent to that in \citet{holden2012self}.}
\begin{equation}
\begin{array}{rl}
LNP(x; \mu, \sigma, \alpha, \rho_P) = & 
\begin{array}{ccl}
\frac{\rho_<}{C\sqrt{2\pi}\sigma x} e^{-\frac{(\ln x - \mu)^2}{2\sigma^2}} & , & x \le x_P \\
\frac{\rho_>}{(1-C)\sqrt{2\pi}\sigma x} e^{-\frac{(\ln x - \mu)^2}{2\sigma^2}} + \frac{\rho_P \alpha x_{P}^{\alpha}}{x^{\alpha+1}} & , & x \ge x_P \\
\end{array}
\end{array}
\end{equation}
Here $C=\frac{1}{2}\left(1+Erf\left(\frac{\alpha \sigma}{\sqrt{2}} \right) \right)$, 
$\alpha$, $\sigma$ and $\rho_P$ are the shape parameters and $x_P$ is the scale
parameter. The other three parameters, $\rho_<$, $\rho_>$ and $\mu$ are determined from the
condition of normalizability of LNP ($\rho_< + \rho_> + \rho_P = 1$) and continuity of LNP and its
derivative at $x_P$ ; in particular, $\mu$ and $x_P$ are related by $x_P = \exp(\mu+\alpha\sigma^2)$.

As mentioned before, the choice of LN front end is motivated by multiplicativity. For
large $x$,
\begin{equation}
\label{LNPTailExponent}
LNP(x; \mu, \sigma, \alpha, \rho_P) \sim x^{-(1+\alpha)}
\end{equation}
As is for GIGa, LNP does not have an upper bound for the tail exponent. However,
unlike S and GIGa, which are infinitely differentiable at all points, LNP's derivatives
above first are discontinuous at $x_P$.

\section{Heavy-Tailed Research Design and Statistics}
\label{HeavyTailedResearchDesignAndStatistics}

Systems that express power-law behavior must be studied with methods that
accommodate more time-dependence and inherent variability than is typically
assumed by linear statistics. The theorems that dictate classical statistical practices
are often stretched to, or beyond, their limits in heavy-tailed applications. For
example, $1/f$ scaling undermines the classical ergodicity assumption, and much
larger sample sizes are required for inference in the context of heavy-tailed systems
than those expressing Gaussian behavior. Extreme observations are so rare in
Gaussian systems that few experiments are designed to discover them, and unusual
observations are reasonably treated as outliers. However, extreme observations are
much more likely in systems that express power-law behavior -- but they may still
be rare. As such it is important to incorporate designs that accommodate much
larger samples than is typical under the assumption of linearity.

Whenever possible, within-subject designs are recommended. If two categories of
behavior are to be contrasted, pairwise yoking of stimuli often yields a design that is
very sensitive for simple distribution contrasts. Each participant is exposed to yoked
target stimuli that are matched on important control variables, but that differ in
terms of the variable of interest. Both contrast distributions can then be comprised
of measurements that originate from the same individuals, on carefully paired target
items that differ only in terms of the variable of interest.

However, many experimental manipulations require between-participants' designs.
In decision studies, for instance, one can use an ideal strategy manipulation \citep{stone1993strategic}. 
This design presents identical positive items but allows one
to vary categories of distractor catch trials. Apparently, the method originated in
lexical decision studies of word recognition, but it could be adapted to other
decision paradigms. This design yields contrast distributions that are comprised of
responses to identical target items from different individuals who completed
conditions with different distractor items. The basic idea behind the manipulation is
to use different classes of distractor items that impact the relative difficulty of the
required decision or discrimination (e.g., the signal-to-noise ratio).

If enough targets can be presented, both the within and between subjects designs
allow one to complete statistical contrasts at the level of both individual participants
and analyses that aggregate across participants. In general, aggregated analyses are
more powerful, and likely more representative of systems that express rare events.
Usually one looks for similar outcomes in aggregated analyses and individual level
analyses.

Similarly, some studies are designed to contrast separate groups of individuals, such
as clinical classification studies, where a group of interest is compared to the control
group. In these studies health impairments may hamper efforts to present enough
trials to individuals to yield stable individual level analyses. Given these
circumstances, aggregated analyses become more important. Rather than fitting
each subject individually, and then averaging the fitting and other parameters over
the group, it is often more expedient to aggregate observations from each
participant into an omnibus distribution that is fit parametrically. There is a
legitimate historical concern in the response time literature that individual and
aggregate analyses have the potential to yield qualitatively different outcomes.
While discrepant results are certainly possible, we now present simulation studies
that test the congruence and validity of individual and aggregate fitting for the
power-law models discussed earlier.

We speculate that response time distributions, with particular parameters,
observed in the context of a particular cognitive activity, requires the support of a
network of particular perceptual, cognitive and neurophysiological process 
\citep{holden2012self, holden2013change, holden2014dyslexic}. At any given point in time,
fluctuations may arise in the organization of this network and its parametric
quantities. This is true both within a task and across group contrasts of interest.
Under the assumption that we are dealing with a network of processes that express
heavy-tailed behavioral measures, individual participant's performances can be
viewed as samples or realizations of random variables that are representative of the
potential states of the process network supporting the cognitive activity of interest.
Given this situation, combining observations across participants may clarify the
potential range of parametric variation that one must expect from a given group or
task, and thus improve the representativeness of any statistical analyses that are
derived from the distribution of observations. Due to the aforementioned
convergence issues, this is especially important for power-law-tail distributions.
Conversely, one expects a far greater parameter variation in subject's power-law-tail 
distributions, especially for small sample sizes.

To contrast the behavior of individual and aggregated parameter estimates we
present the following numerical simulation using GIGa as a candidate distribution:
\begin{enumerate}[(1)]
\item \label{simuFirst} We fit the aggregated RT trials data of each of the two groups with GIGa to
obtain the initial values of its parameters $\alpha_0^i$, $\beta_0^i$, $\gamma_0^i$, $\mu_0^i$, where $i = 1, 2$ is for
control and dyslexic group respectively (for instance, the values in Table \ref{table1} 
are the result of such fits).
\item \label{simuSecond}  We use parameters $\alpha_0^i$, $\beta_0^i$, $\gamma_0^i$, $\mu_0^i$, 
obtained in step (\ref{simuFirst}), to generate the group-size 
variates. These variate "groups" are meant to serve as simulated
counterparts of the RT trials control (C) and dyslexic (D) groups used in step (\ref{simuFirst}).
\item \label{simuThird}  We use the bootstrap procedure \citep{efron1994introduction} on the group
variates from step (\ref{simuSecond}) to obtain the bootstrap distribution, mean, standard deviation (SD) and
confidence interval (CI) for the parameters $\alpha^i$, $\beta^i$, $\gamma^i$, $\mu^i$ of the group variates.
There are other ways to generate the variability of parameters -- we chose to
use bootstrap in order to parallel the procedure of Sec. \ref{ApplicationToDyslexia} on RT trials group
data.
\item \label{simuFourth}  We then generate random variates of the individual-subject size, where the
parameters of the candidate distribution are taken at random from a normal
distribution, whose mean is the real data parameter from step (\ref{simuFirst}) and SD is
obtained in step (\ref{simuThird}), mimicking the variability of the parameters mentioned
above. These subject-size distributions -- variate "subjects" -- are individually
fitted and we obtain the mean, SD and range of the parameters of these.
\item \label{simuFifth}  We then compare the variate "groups" and variate "subjects" to the actual RT
trials data groups and subjects.
\end{enumerate}

As mentioned above, there are alternative methods of generating parameter
variability, described in step (\ref{simuThird}). For instance, one could resample the raw data
distributions or use the Jackknife procedure \citep{efron1994introduction}. In this
regard, it should be noted that the bootstrap distributions of parameters of
simulated variates are narrower than those of RT trials data, as seen in Figures \ref{figure1}
and \ref{figure2} (see below). The main conclusion, however, is that a distribution of parameters of 
aggregated variates are narrower than those derived from individual's data, be it the
RT trials data or their simulated counterparts. This illustrates a potential benefit of
aggregation of individual observations. This could be particularly useful when
individual observations exert undue influence on parameter estimates, as when
sample sizes are small.

Here we implement this approach for the Arithmetic study that contrasted
children with and without dyslexia (see Sec. \ref{Rescaling} below and \citep{holden2014dyslexic}).
Participants pressed one button to indicate that displayed simple addition sums,
together with an answer, were correct (e.g., $3+6=9$) and another button if incorrect
(e.g., $4+3=2$). The answers to all the sums were always below $10$.

Using GIGa as a candidate distribution,
\footnote{As was mentioned above, GIGa corresponds to the generalized Bouchaud-M$\acute{e}$zard
network model \citep{bouchaud2000wealth, ma2013distribution}. Without ascertaining
any direct relationship of the latter to the actual neural network, this is in line with
the hypothesis of a close correspondence between the observed RT distribution and
the underlying cognitive neurophysiology of a particular task in the neural network.}
$20$ variate "Psudo-subjects" of $560$ points each were generated, per steps (1)-(4) above, similar to the actual trials. As an
example, we use two measures of variability -- the tail exponent is computed from
the GIGa parameters as $(\alpha\gamma+1)$ in Eq. (\ref{GIGaTailExponent}) and the half width (HW) of the
distribution, defined as the width of the distribution along the line drawn at half
height of the PDF's maximum, that is, modal value of PDF (MPDF) -- see Sec. \ref{ApplicationToDyslexia}
below -- both of which depend on the shape parameters. The results are summarized in Figures \ref{figure1}
and \ref{figure2} and Tables \ref{table1}--\ref{table4}, where "C" stands for control and "D"  for dyslexic.

\begin{figure}[!htbp]
\centering
\begin{tabular}{cc}
\includegraphics[width = \myFigureWidth \textwidth]{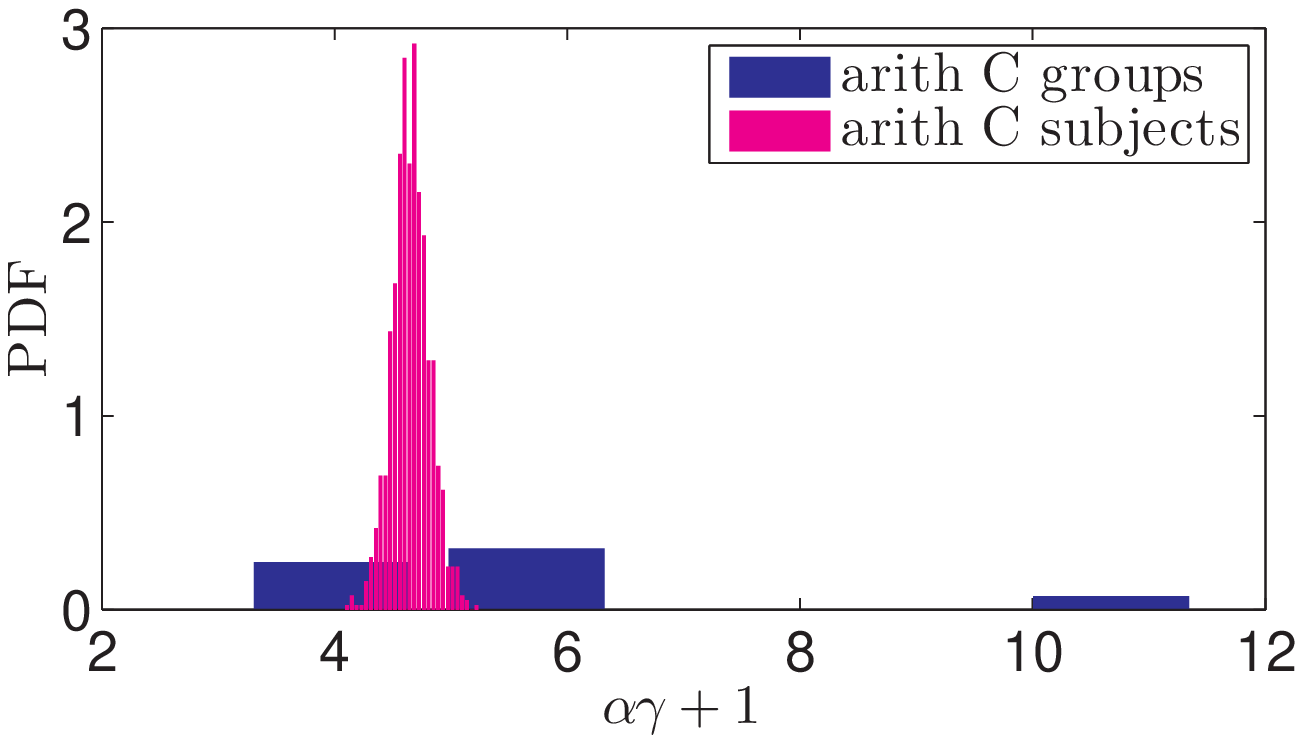} & 
\includegraphics[width = \myFigureWidth \textwidth]{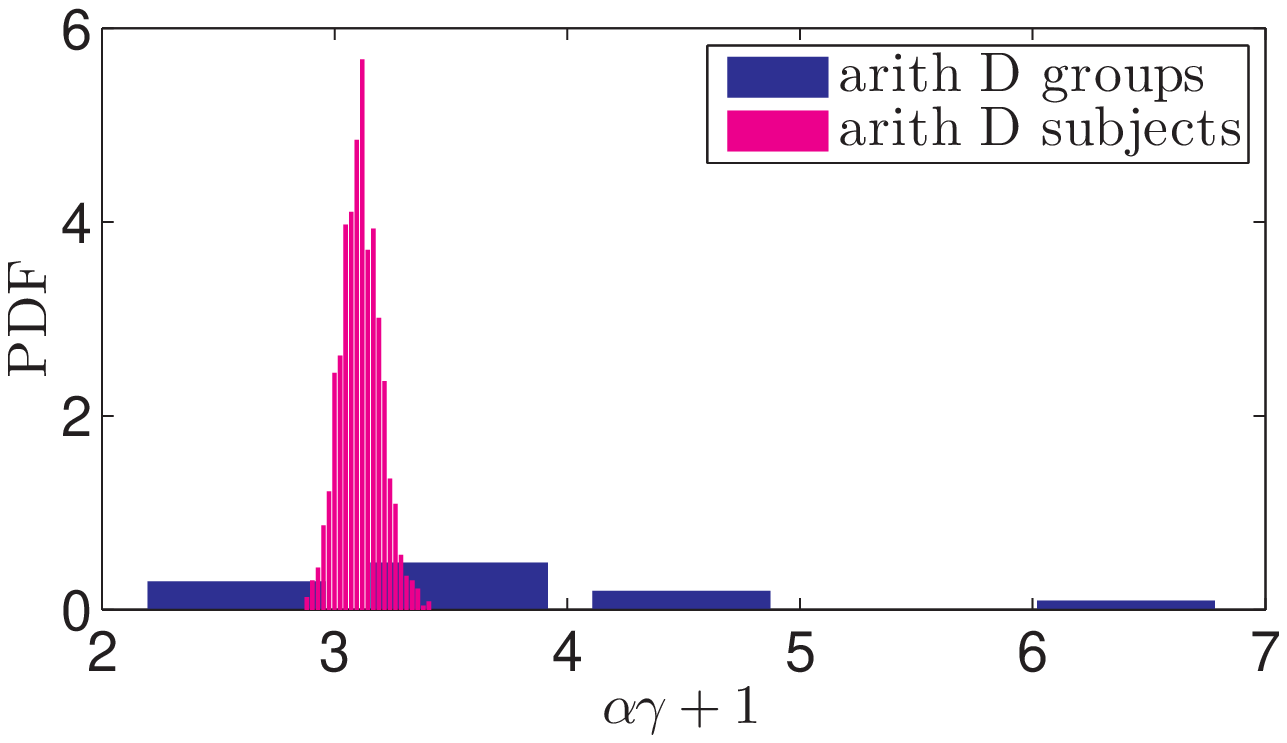} \\
\includegraphics[width = \myFigureWidth \textwidth]{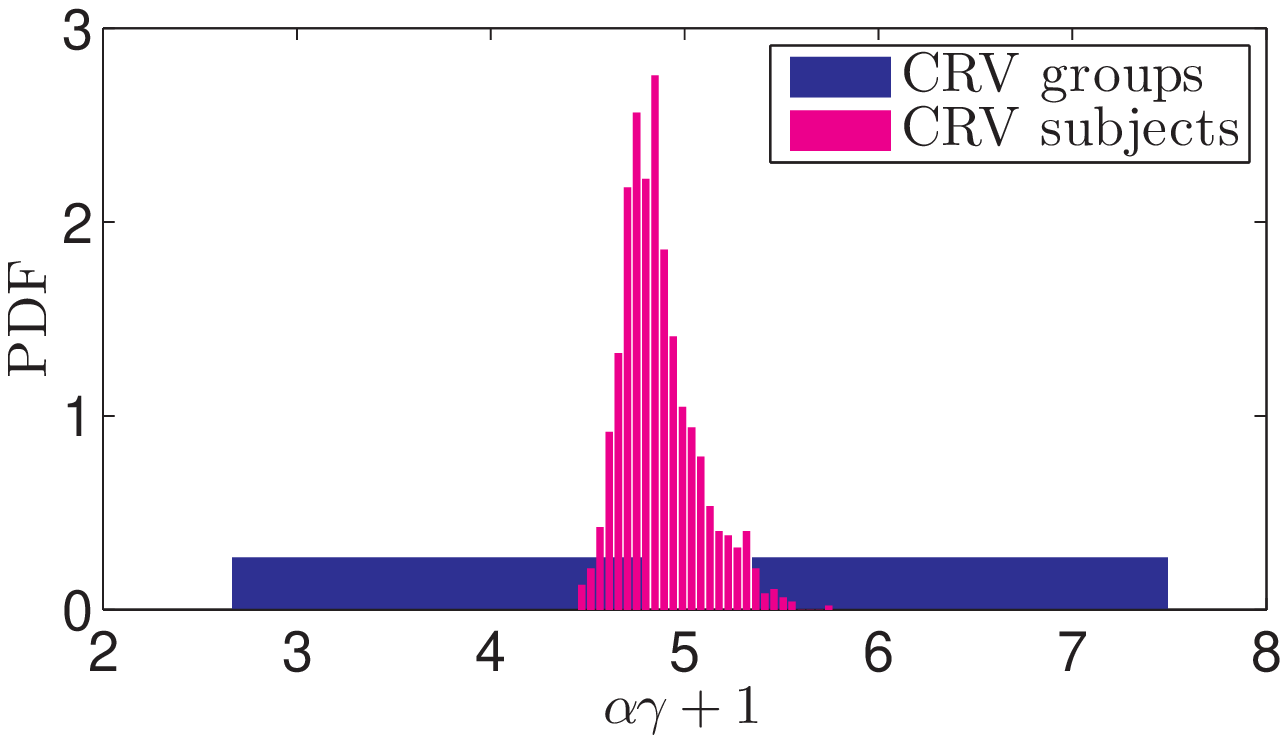} & 
\includegraphics[width = \myFigureWidth \textwidth]{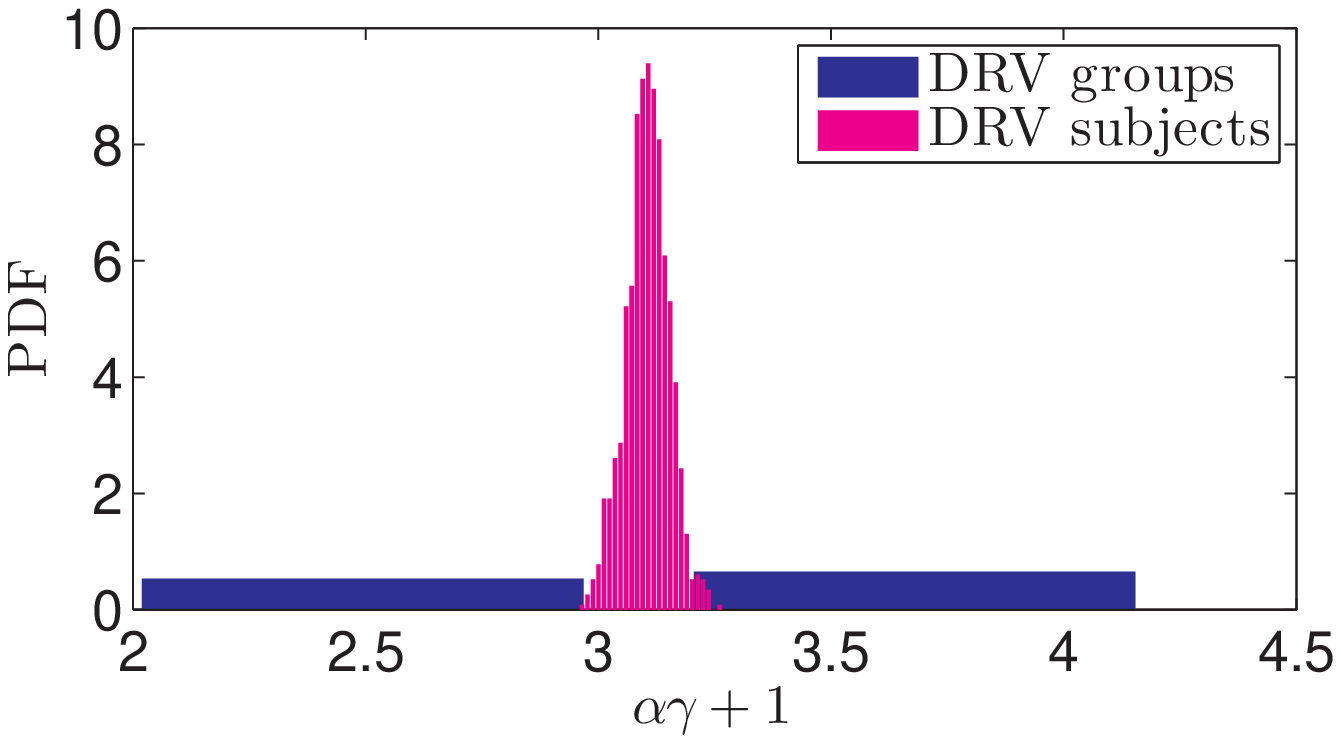} \\
\end{tabular}
\caption{Tail exponent ($\alpha\gamma+1$). Top row: Histograms of parameters derived from
fits of individual subjects (blue bars) and from fits of group (aggregate) data (magenta 
bars), obtained in RT trials for the control and dyslexic groups respectively. Bottom
row: Histograms of parameters derived from fits of variate "subjects" and variate
"groups" -- counterparts of the individual and group RT data -- obtained in steps (\ref{simuThird})
and (\ref{simuFourth}) of the random variate simulation described in text.}
\label{figure1}
\end{figure}

Tables \ref{table1} and \ref{table3} contain the results for actual trials data and the statistics for the
combined group data are calculated using the bootstrap. In Tables \ref{table2} and \ref{table4}, CRV and
DRV are control group and dyslexic group variates respectively; mean, SD and CI for the
variate "groups" are obtained in step (\ref{simuThird}); mean, SD and range for the variate
"subjects" are obtained in step (\ref{simuFourth}). CI in Tables \ref{table1}-\ref{table4} are set to the $95\%$ confidence
level.

\begin{table}[!htbp]
\centering
\caption{Tail Exponent $\alpha\gamma+1$}
\label{table1}
\begin{tabular}{cccrlccrl} 
\hline
 & \multicolumn{4}{c}{Actual Group} & \multicolumn{4}{c}{Actual Subjects} \\
\hline
 & mean & SD & \multicolumn{2}{c}{CI} & mean & SD & \multicolumn{2}{c}{Range} \\
\hline
C & 4.656 & 0.161 & (4.469, & 5.114) & 5.383 & 2.240 & (3.140, & 11.511) \\
D & 3.112 & 0.087 & (2.948, & 3.291) & 3.801 & 1.241 & (2.102, & \, 6.879) \\
\hline
\end{tabular}
\end{table}

\begin{table}[!htbp]
\centering
\caption{Variate Tail Exponent $\alpha\gamma+1$}
\label{table2}
\begin{tabular}{cccrlccrl} 
\hline
 & \multicolumn{4}{c}{"Pseudo-Group" Variate} & \multicolumn{4}{c}{"Pseudo-Subjects" Variate} \\
\hline
 & mean & SD & \multicolumn{2}{c}{CI} & mean & SD & \multicolumn{2}{c}{Range} \\
\hline
CRV & 4.868 & 0.198 & (4.479, & 5.135) & 4.941 & 1.560 & (2.398, & 7.760) \\
DRV & 3.106 & 0.046 & (3.055, & 3.235) & 3.130 & 0.712 & (1.901, & 4.272) \\
\hline
\end{tabular}
\end{table}

Comparing the dispersion of the synthetic individual fits to that of the 
all-in-one fits recommends an all-in-one approach. Given the potential noisiness of
performance data from children, and that each group, control and dyslexic,
contribute only about $10^4$ combined data points, the quantitative agreement seems
to be quite good as well.

Likewise, contrasts of the confidence intervals for the synthetic group variables and
actual group data with the ranges for subjects', demonstrates that fitting of
individual subjects, with subsequently averaged parameters, may be not sufficiently
accurate to distinguish rescaled distributions (see Sec. \ref{Rescaling} and \ref{ApplicationToDyslexia}) from those of
different shape and, by proxy, distinct underlying relations among cognitive and
neurophysiological processes.

\begin{figure}[!htbp]
\centering
\begin{tabular}{cc}
\includegraphics[width = \myFigureWidth \textwidth]{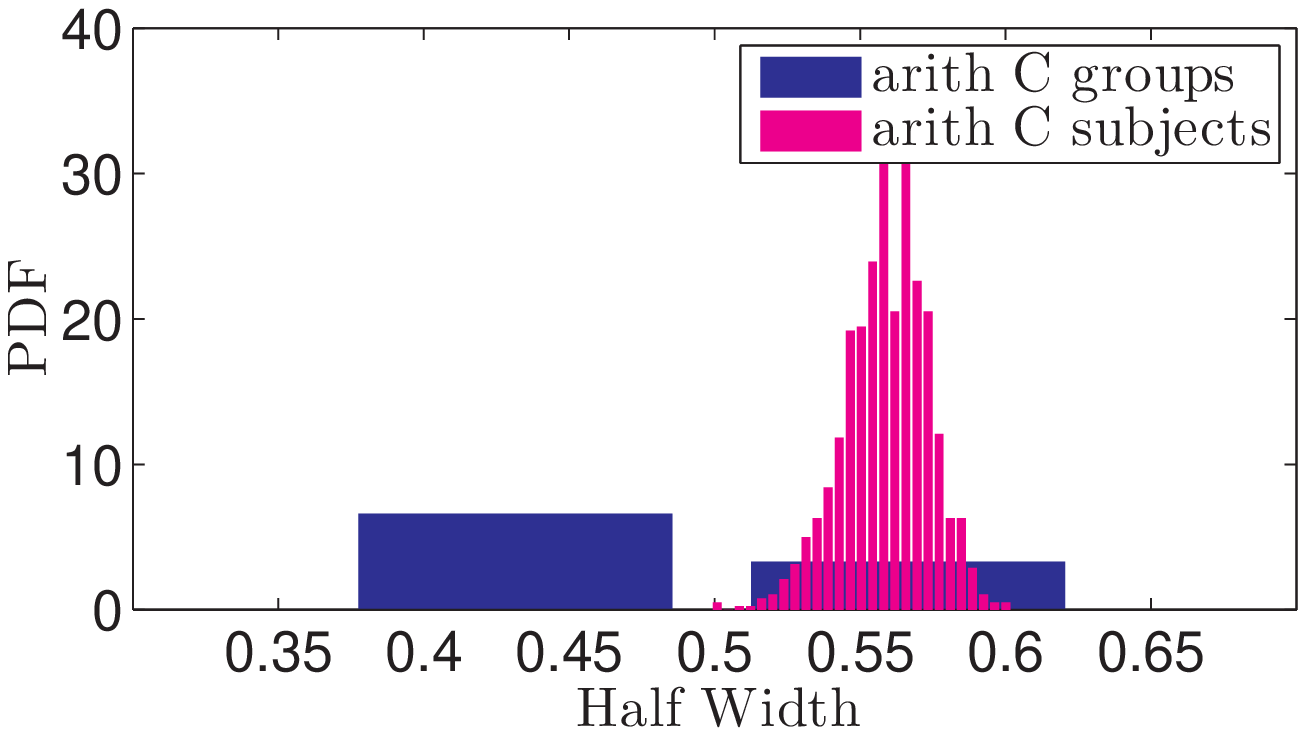} & 
\includegraphics[width = \myFigureWidth \textwidth]{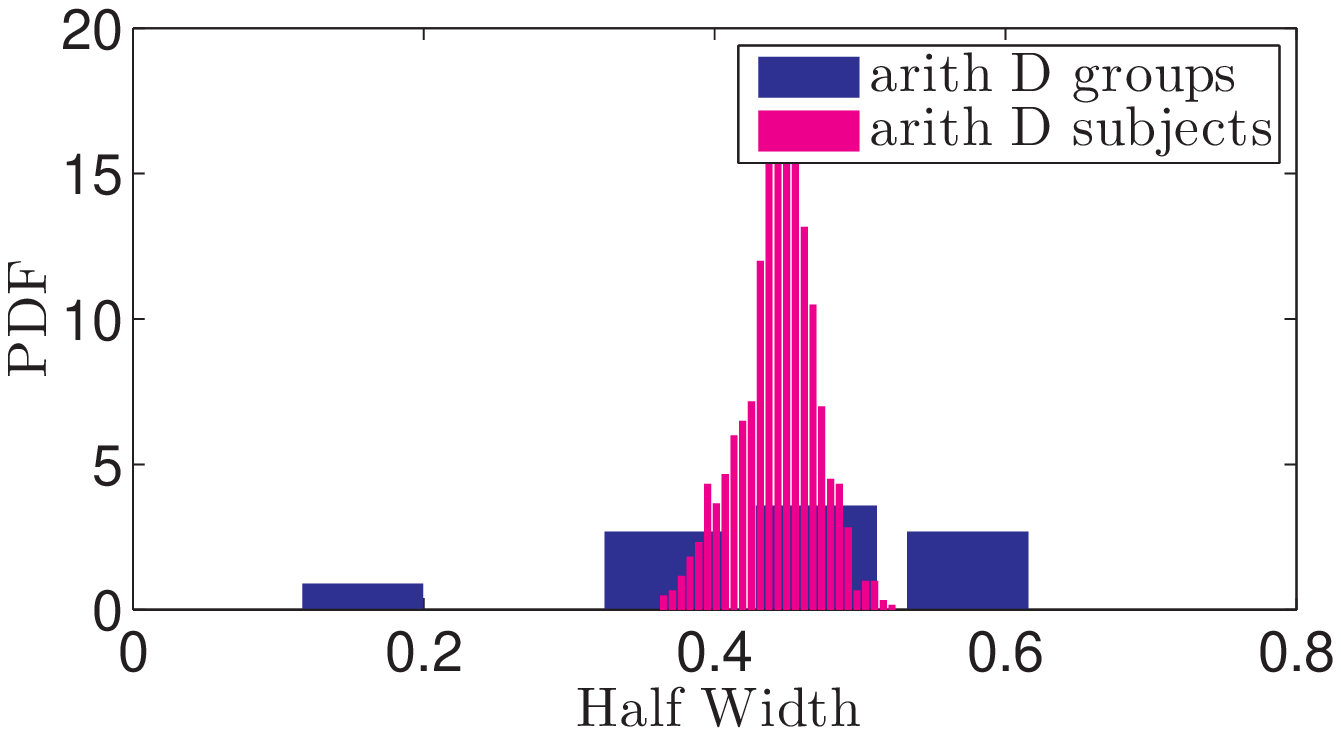} \\
\includegraphics[width = \myFigureWidth \textwidth]{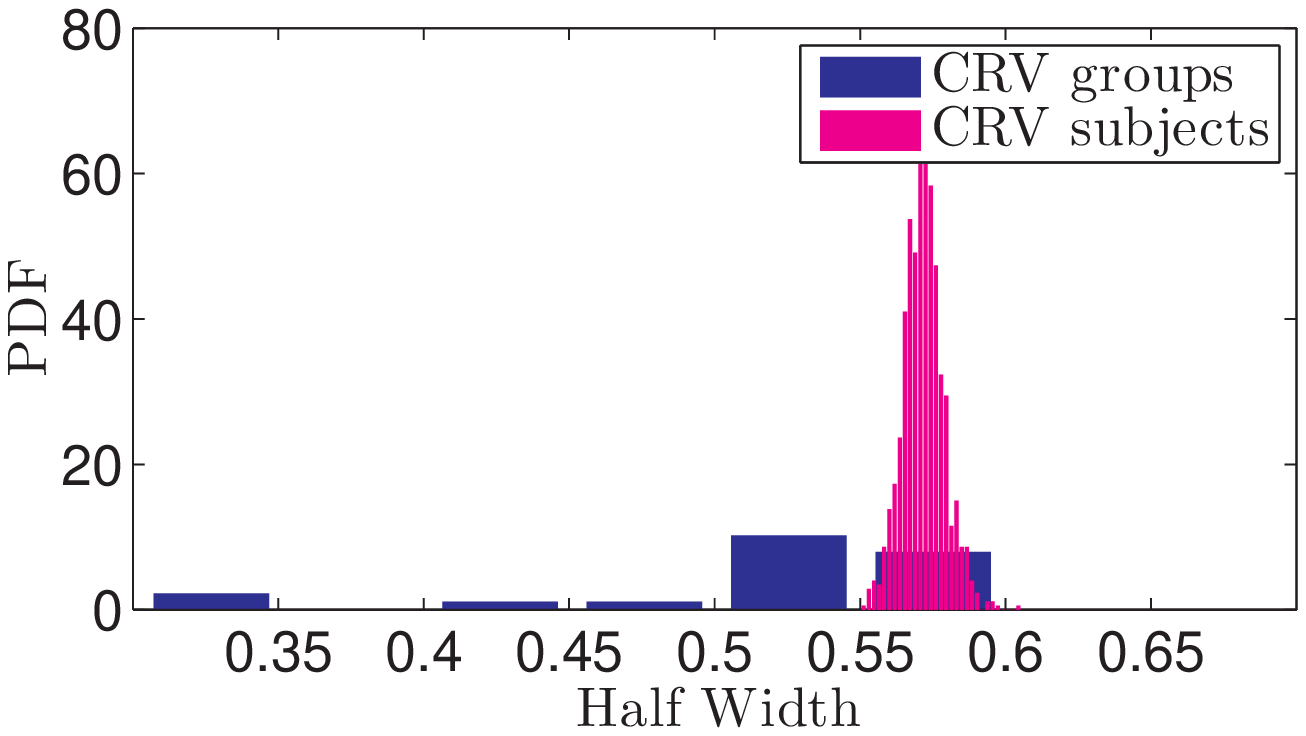} & 
\includegraphics[width = \myFigureWidth \textwidth]{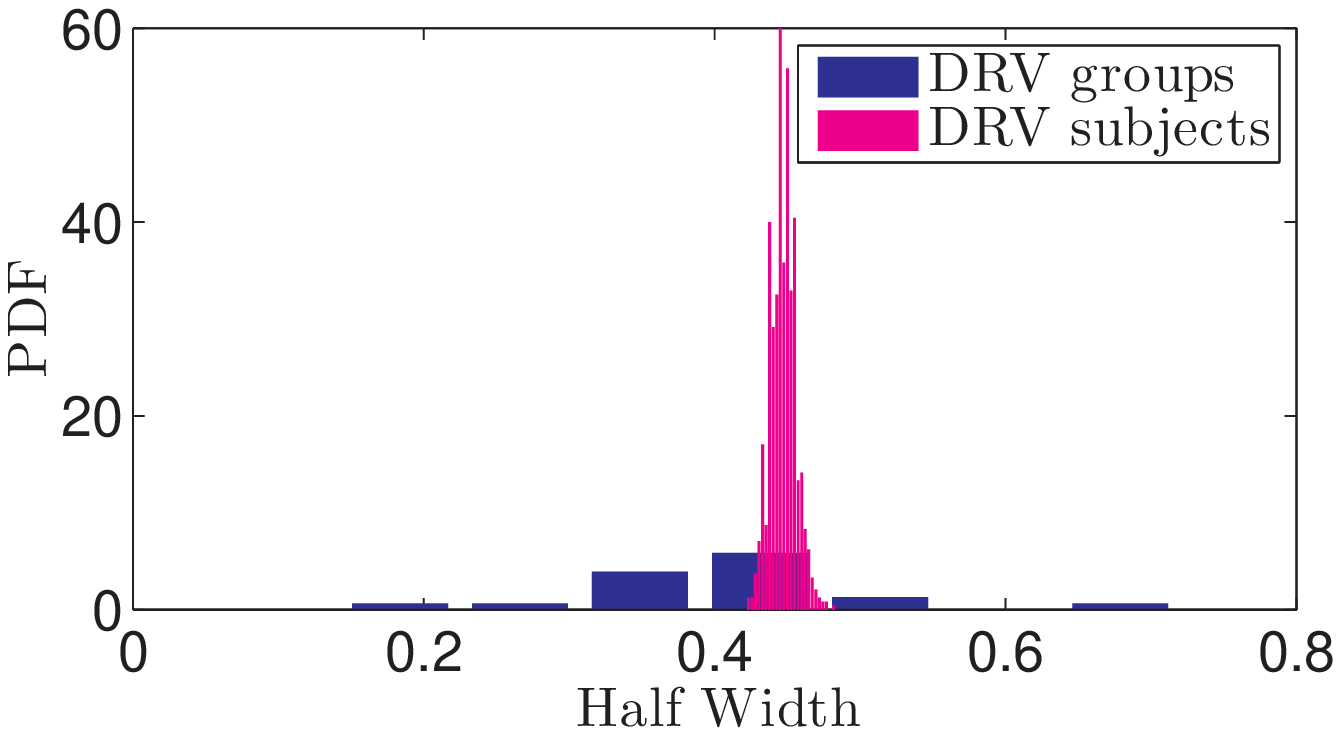} \\
\end{tabular}
\caption{Half Width histograms. Top row: Histograms of parameters derived from
fits of individual subjects (blue bars) and from fits of group (aggregate) data (magenta 
bars), obtained in RT trials for the control and dyslexic groups respectively. Bottom
row: Histograms of parameters derived from fits of variate "pseudo-subjects" and
variate "pseudo-groups" -- counterparts of the individual and group RT data --
obtained in steps (\ref{simuThird}) and (\ref{simuFourth}) of the random variate simulation described in text.}
\label{figure2}
\end{figure}

\begin{table}[!htbp]
\centering
\caption{Half-Width (HW)}
\label{table3}
\begin{tabular}{cccrlccrl} 
\hline
 & \multicolumn{4}{c}{Actual Group} & \multicolumn{4}{c}{Actual Subjects} \\
\hline
 & mean & SD & \multicolumn{2}{c}{CI} & mean & SD & \multicolumn{2}{c}{Range} \\
\hline
C & 0.559 & 0.015 & (0.543, & 0.599) & 0.463 & 0.087 & (0.364, & 0.634) \\
D & 0.444 & 0.026 & (0.380, & 0.487) & 0.447 & 0.140 & (0.106, & 0.626) \\
\hline
\end{tabular}
\end{table}

\begin{table}[!htbp]
\centering
\caption{Variate Half-Width (HW)}
\label{table4}
\begin{tabular}{cccrlccrl} 
\hline
 & \multicolumn{4}{c}{"Pseudo-Group" Variate} & \multicolumn{4}{c}{"Pseudo-Subjects" Variate} \\
\hline
 & mean & SD & \multicolumn{2}{c}{CI} & mean & SD & \multicolumn{2}{c}{Range} \\
\hline
CRV & 0.572 & 0.007 & (0.559, & 0.587) & 0.515 & 0.079 & (0.302, & 0.600) \\
DRV & 0.447 & 0.009 & (0.433, & 0.469) & 0.405 & 0.112 & (0.142, & 0.720) \\
\hline
\end{tabular}
\end{table}

Next, we discuss several interpretable differences that can appear in the parameters
of heavy-tailed distributions arising from systems that are governed by interaction-dominant 
dynamics \citep{holden2009dispersion, van2013fractal}. Finally, we illustrate
the technique on data from a recently published dyslexia study \citep{holden2014dyslexic}.

\section{Rescaling}
\label{Rescaling}

Suppose one collects RT data, in an identical experimental setup, from two distinct
groups of participants -- for instance, children with and without dyslexia. Is it
possible to relate the underlying relationships among cognitive and
neurophysiological systems to the shape of the response time PDF? If the aggregate
distributions of the two groups happen to be the rescaled versions of each other, we
posit that the supporting networks of perceptual, cognitive and neurophysiology
processes is functioning in a qualitatively similar manner, as the only difference
entailed in rescaling is time dilation or time contraction. 
By contrast, substantial shape differences imply the functional organization of the
supporting perceptual, cognitive and neurophysiological systems is different. So if 
\begin{equation}
\label{scaleEquation}
PDF_2(c \cdot x) = \frac{1}{c}PDF_1(x)
\end{equation}
where
\begin{equation}
\label{scaleConstant}
c = \frac{mean_2}{mean_1}
\end{equation}
then it is indeed the same, with one group simply being proportionally faster (or
slower) than the other. In this case the two PDF can be transformed into one by
scaling back to enforce the same parameters (see below).

Ordinarily, the rescaling of an analytically defined PDF is achieved via rescaling of
the scale parameter of a distribution. A familiar example of a scale parameter is the
standard deviation $\sigma$ of the normal distribution 
\begin{equation}
\label{normalDistribution}
N(x; \mu, \sigma) = \frac{1}{\sqrt{2\pi}\sigma} e^{-\frac{(x-\mu)^2}{2\sigma^2}}
\end{equation}
where the mean $\mu$ is a familiar example of a location parameter. Indeed, subjecting
Eq. (\ref{normalDistribution}) to $\sigma\rightarrow c\cdot \sigma$ 
(and simultaneously $\mu \rightarrow c\cdot \mu$) produces a set of two PDF related by Eqs. (\ref{scaleEquation})
and (\ref{scaleConstant}). Rescaling the variable, $x \rightarrow c\cdot x$, in $PDF_2$ will bring the two distribution back
into one via $c \cdot PDF_2(c\cdot x) = PDF_1(x)$. On the other hand, for a lognormal distribution
(LN),
\begin{equation}
LN(x; \mu, \sigma) = \frac{1}{\sqrt{2\pi}\sigma x} e^{-\frac{(\ln x - \mu)^2}{2\sigma^2}}
\end{equation}
$\sigma$ is a shape parameter and $\mu$ -- or rather $\exp(\mu)$ -- is a scale parameter. The LN is a
heavy-tail distribution, which enjoys wide utility across many fields
\citep{log2001log}. It is used here as the front end of the LNP distribution. Just as
N and S distributions are a paradigm for additivity via the central and generalized
central limit theorem, LN is a paradigm for multiplicativity (additivity of the log).

In those cases where we can identify the scale and the shape parameters, our initial
statement can be reformulated as follows: when the two RT distributions can be
transformed into each other via rescaling of the scale parameter -- without affecting
the shape parameters -- the underlying neurophysiology behind these distributions
is effectively identical, while the substantive difference in the shape parameters is
indicative of an alternate functional organization.

\begin{figure}[!htbp]
\centering
\includegraphics[width = \myFigureWidth \textwidth]{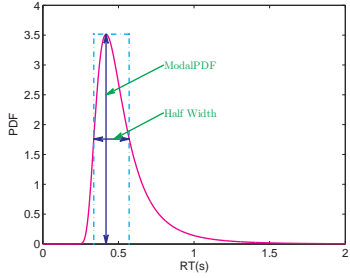}
\caption{Depicts the relationship between a PDF's modal value and its half width.}
\label{figure3}
\end{figure}

We use two shape-related markers to gauge variability: the half width
(HW) of the distribution and the exponent of the power-law tail of PDF. HW is
defined as the width of the distribution along the line drawn at half height of the
PDF's maximum, the modal value of PDF (MPDF) as shown in Figure \ref{figure3}. The
product of HW and MPDF defines the area of the dashed box in Figure \ref{figure3}. For
sufficiently large samples, this product typically lies in a narrow range centered
around $0.8$ and approximates the fraction of PDF that is not in the tail.

We use HW, instead of SD, because PDF's fat tail heavily influences the value of the
latter. Both HW and power-law exponent depend on at least one shape parameter.
Unlike the simple dependence of the latter (see Eqs. (\ref{GIGaTailExponent}) and (\ref{LNPTailExponent})), the parametric
dependence of HW can be quite complex. For GIGa, for instance, the expressions for
HW and MPDF for $\mu = 0$ are as follows:
\begin{equation}
\begin{array}{rcl}
HW & = & \beta \left\{ \left[ -\frac{\alpha\gamma+1}{\gamma}W_0\left( 
-\frac{2^{-\frac{\gamma}{\alpha\gamma+1}} \left( \frac{\alpha\gamma+1}{e \gamma} \right)^{\left( 
\frac{\alpha\gamma+1}{\gamma} \right)}}{\alpha+\gamma^{-1}} \right)\right]^{-\frac{1}{\gamma}} \right.\\
 & & \quad \left. - \left[ -\frac{\alpha\gamma+1}{\gamma}W_{-1}\left(-\frac{2^{-\frac{\gamma}{\alpha\gamma+1}}
\left(\frac{\alpha\gamma+1}{e\gamma} \right)^{\left(\frac{\alpha\gamma+1}{\gamma} \right)}}{\alpha+\gamma^{-1}} 
\right) \right]^{-\frac{1}{\gamma}} \right\}
\end{array}
\end{equation}

\begin{equation}
MPDF = \frac{\gamma \left(\frac{\alpha\gamma+1}{e\gamma}\right)^{
\left(\frac{\alpha\gamma+1}{\gamma}\right)}}{\beta \Gamma(\alpha)}
\end{equation}
where $W_0$ and $W_{-1}$ are the two branches of the Lambert W function and $\Gamma$ is the
gamma function.

\section{Application to Dyslexia}
\label{ApplicationToDyslexia}

We now use GIGa and LNP distributions to decide whether the difference between
the dyslexic and control groups, described in detail in \citep{holden2014dyslexic}, reduces
to rescaling or implies a more significant neurophysiological difference. Holden et al.
contrasted the performance of children with dyslexia and matched age-appropriate
readers in four tasks with varied reading related requirements. In the above
reference, each subject was fitted with an LNP and parameters thus obtained were
averaged between subjects, which is a standard technique currently in use. Based on
the latter, the conclusion was drawn that the difference between the control and
dyslexic groups is consistent with rescaling in the reading intensive task. This
method averaged parameters of individual fits, and was not sensitive enough to
detect systematic between-group differences in the non-reading tasks.

Here we use a less-traditional averaging procedure, based on aggregation analysis of
Sec. \ref{HeavyTailedResearchDesignAndStatistics}, that reveals additional information. 
Namely, we combine the RT of the entire control group into a single sample and do the 
same for the dyslexic group. We believe that such an averaging procedure may be more 
sensitive to subtle trends in clinical studies, as discussed in terms of random variates in 
Sec. \ref{HeavyTailedResearchDesignAndStatistics}.

In particular, we analyze the flanker and the arithmetic tasks \citep{holden2014dyslexic}. 
\footnote{The full study \citep[][in preparation]{liuInPreparation} will be reported elsewhere.}
While the former is a better test of a lower-level skill, the latter emphasizes the
higher-level cognitive activity. We argue that while the flanker is consistent with the
rescaling, arithmetic task is not. Once we fitted the distributions with GIGa and LNP,
we used a bootstrap procedure \citep{efron1994introduction} to determine the
confidence intervals and distributions of the parameters.


\subsection{Flanker}

\begin{figure}[!htbp]
\centering
\begin{tabular}{cc}
\includegraphics[width = \myFigureWidth \textwidth]{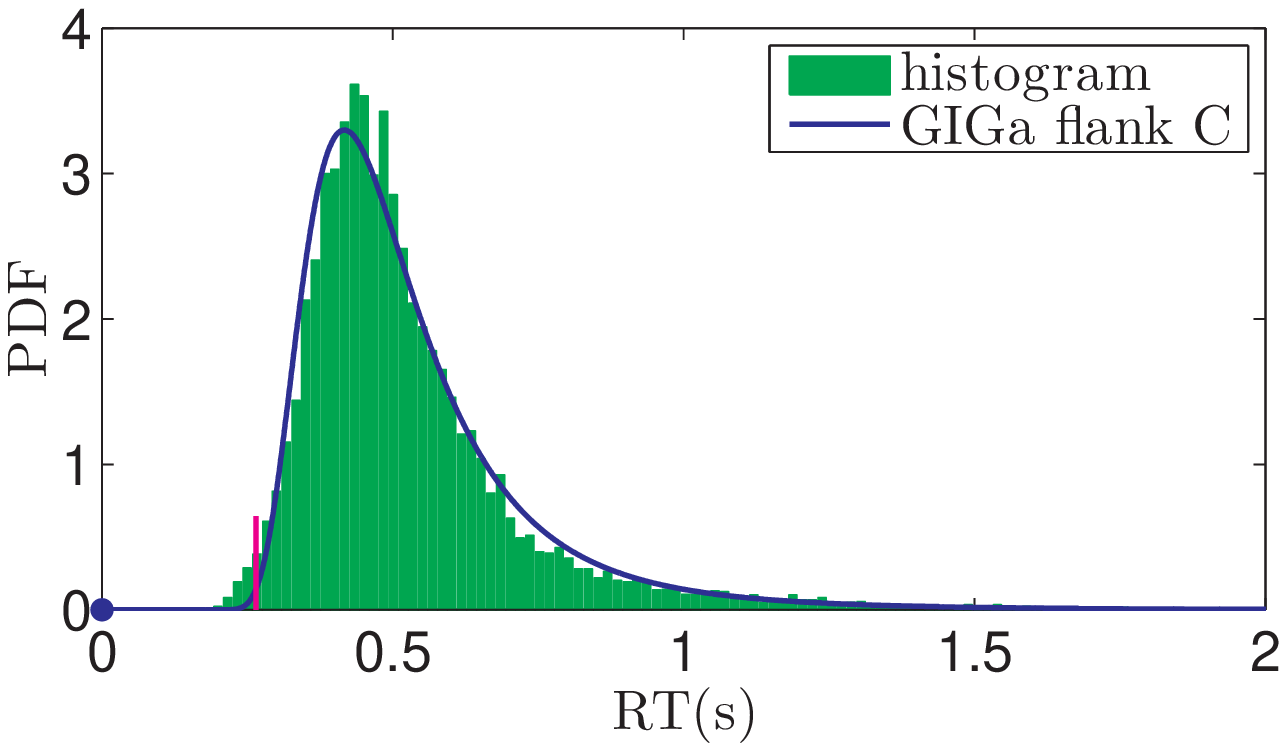} & 
\includegraphics[width = \myFigureWidth \textwidth]{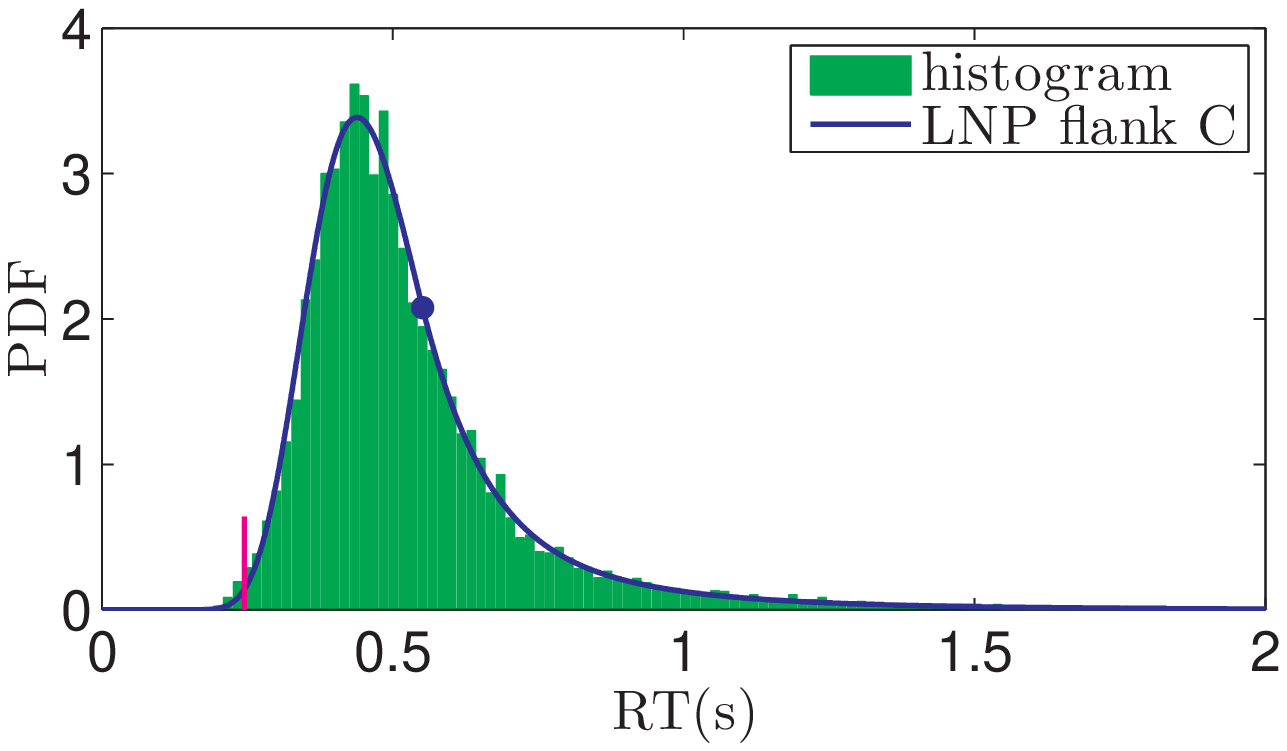} \\
\includegraphics[width = \myFigureWidth \textwidth]{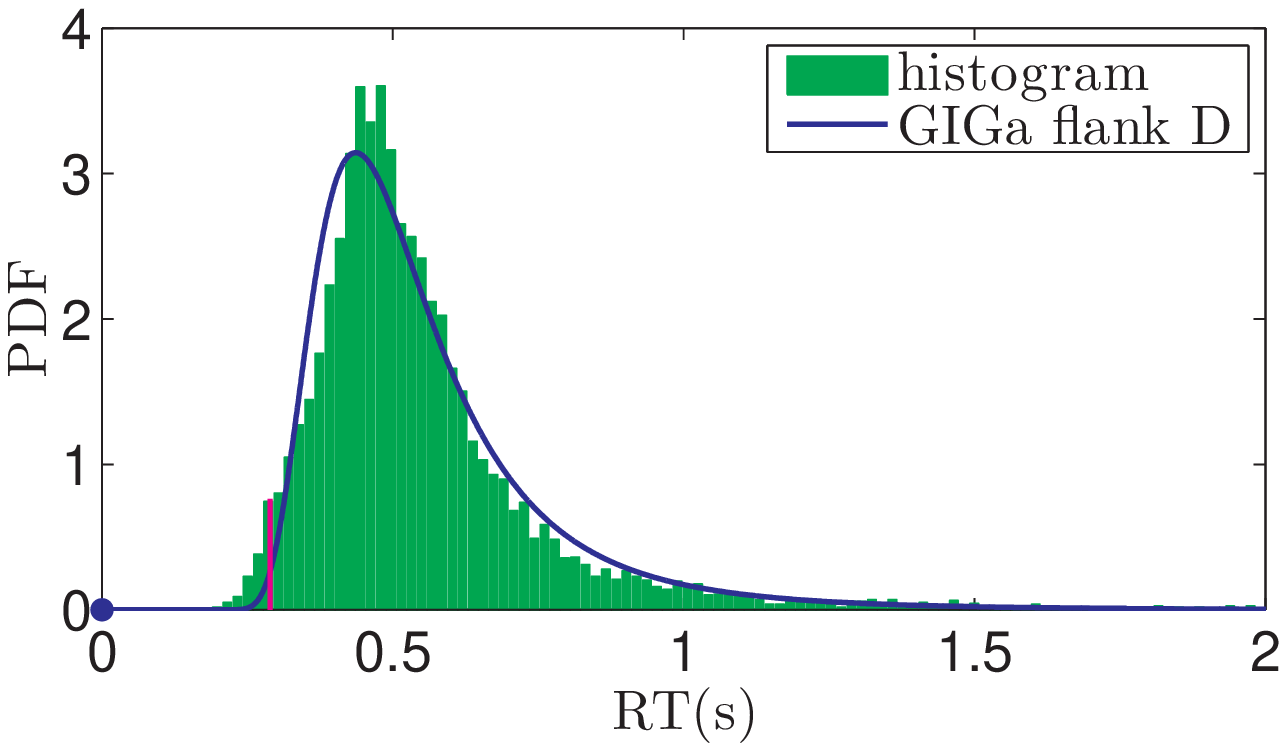} & 
\includegraphics[width = \myFigureWidth \textwidth]{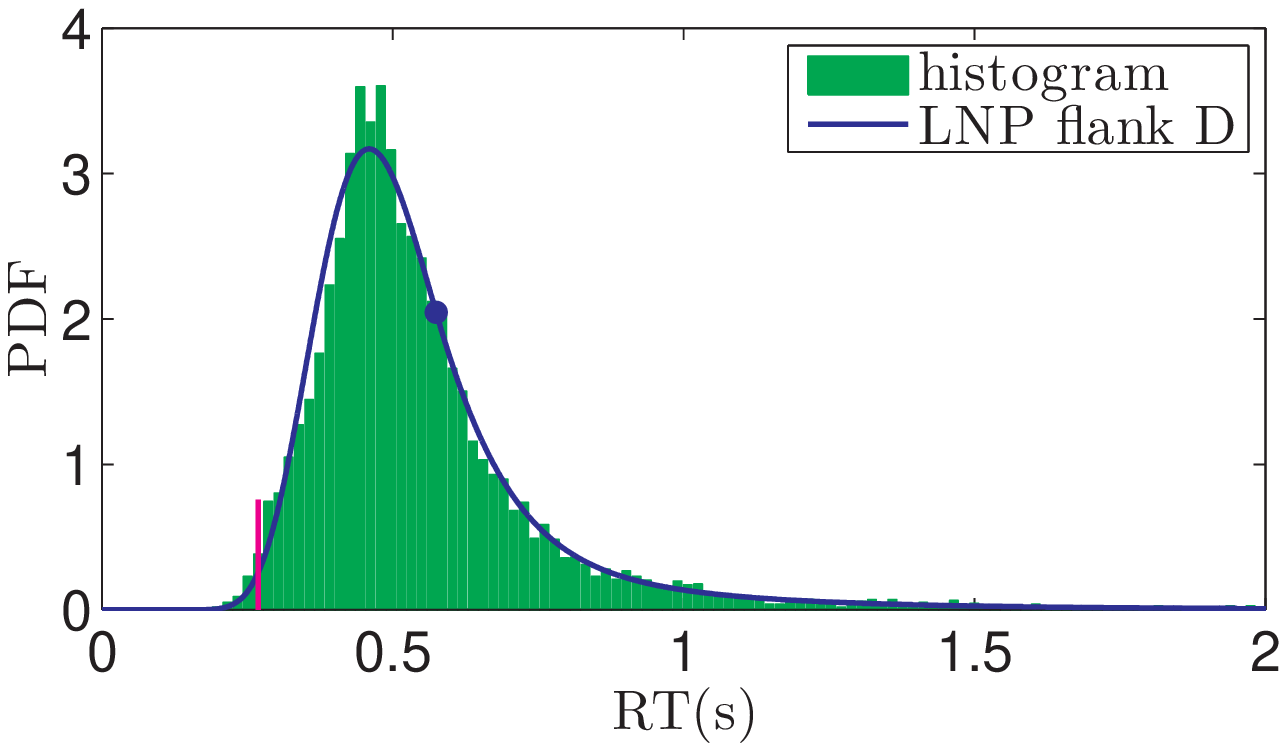} \\
\includegraphics[width = \myFigureWidth \textwidth]{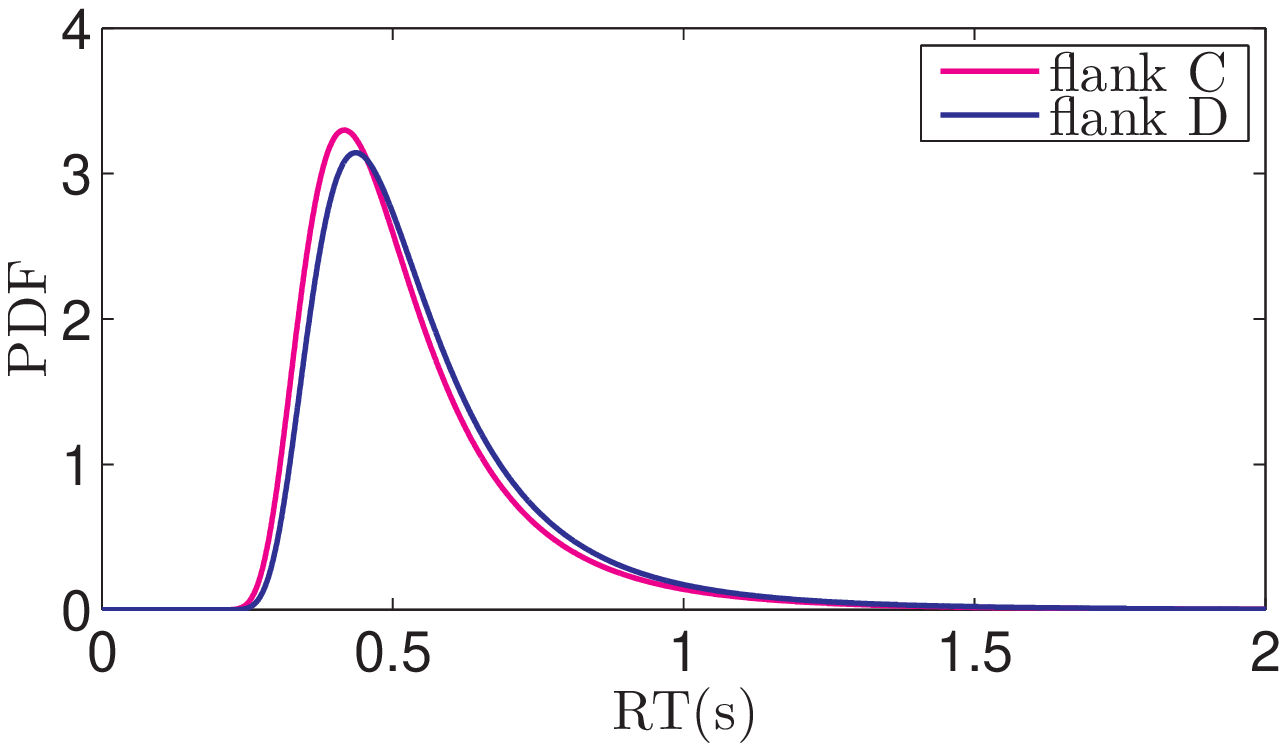} & 
\includegraphics[width = \myFigureWidth \textwidth]{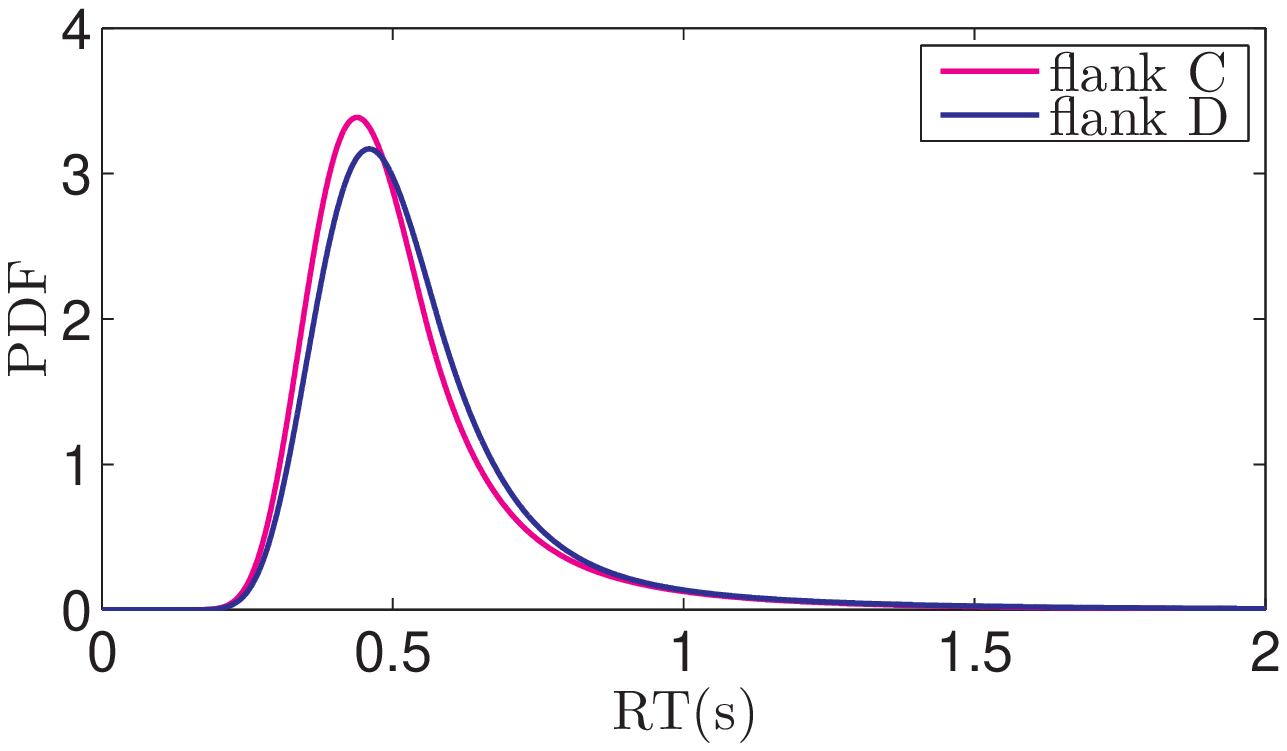} \\
\includegraphics[width = \myFigureWidth \textwidth]{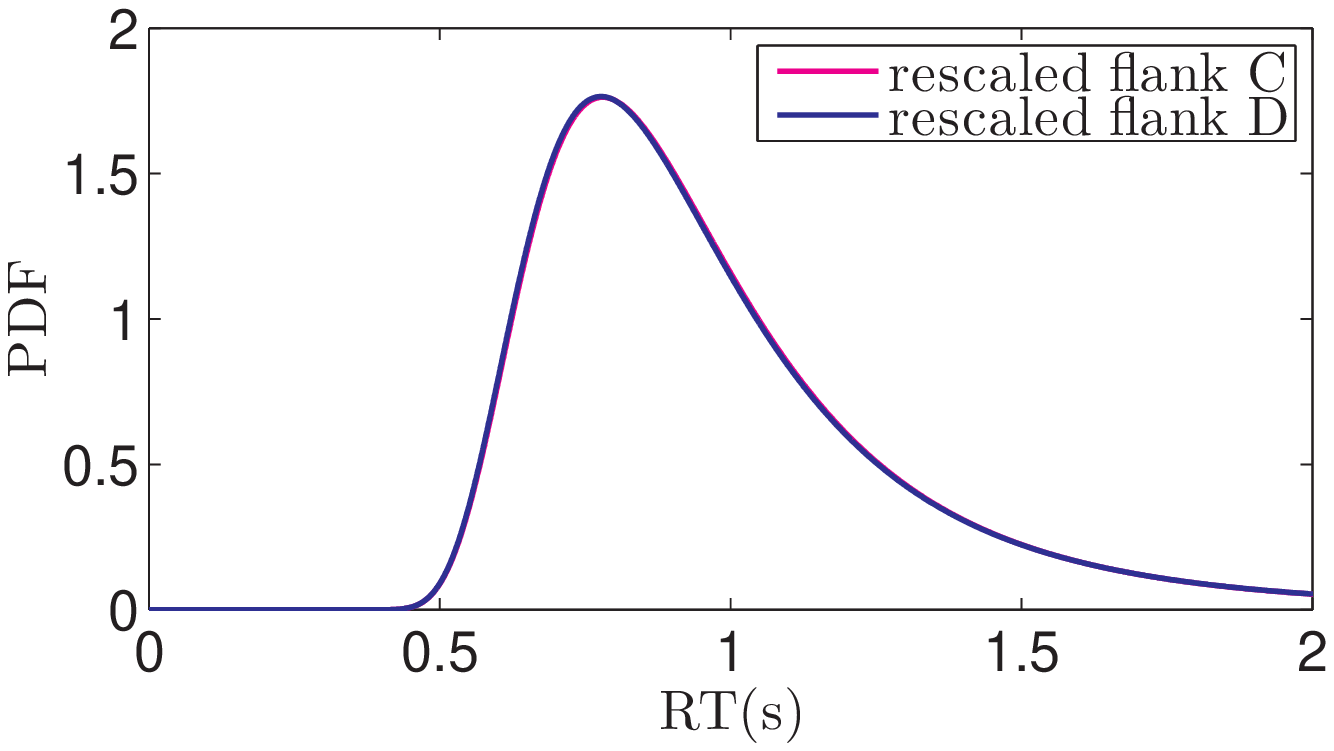} & 
\includegraphics[width = \myFigureWidth \textwidth]{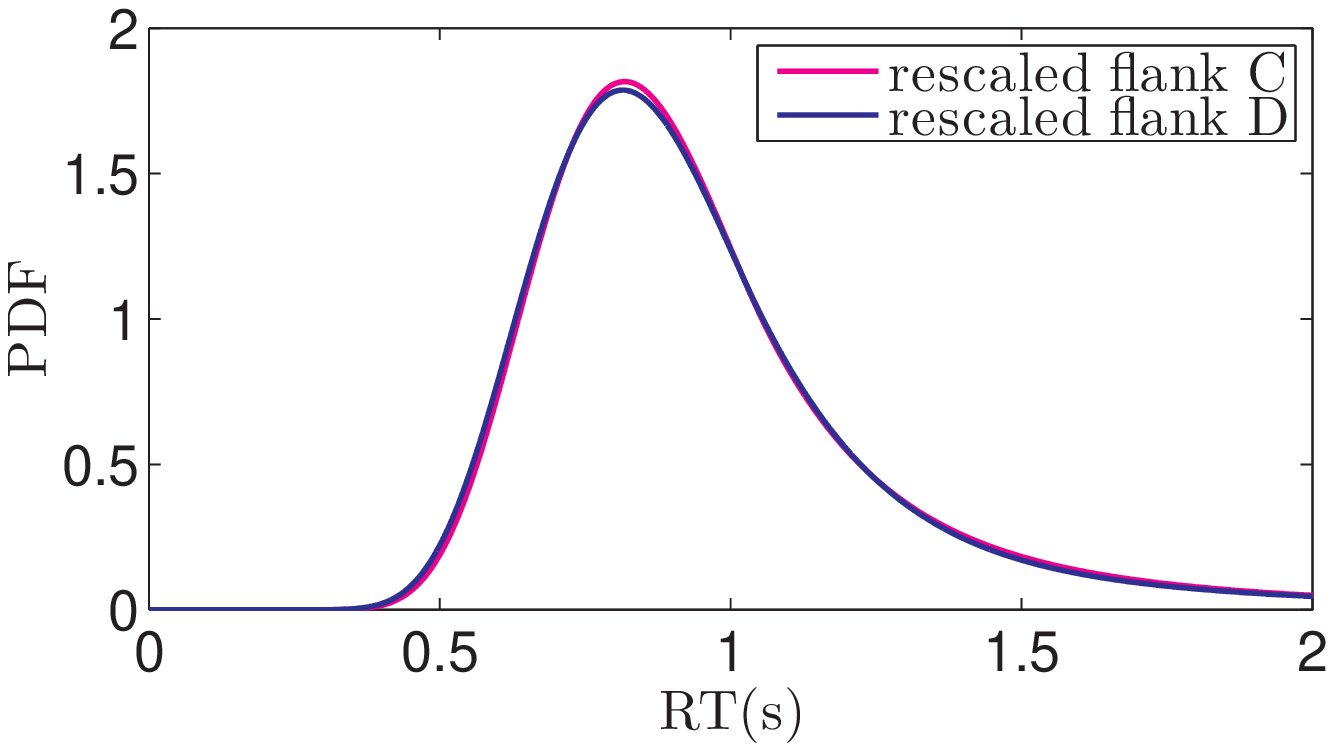} \\
\includegraphics[width = \myFigureWidth \textwidth]{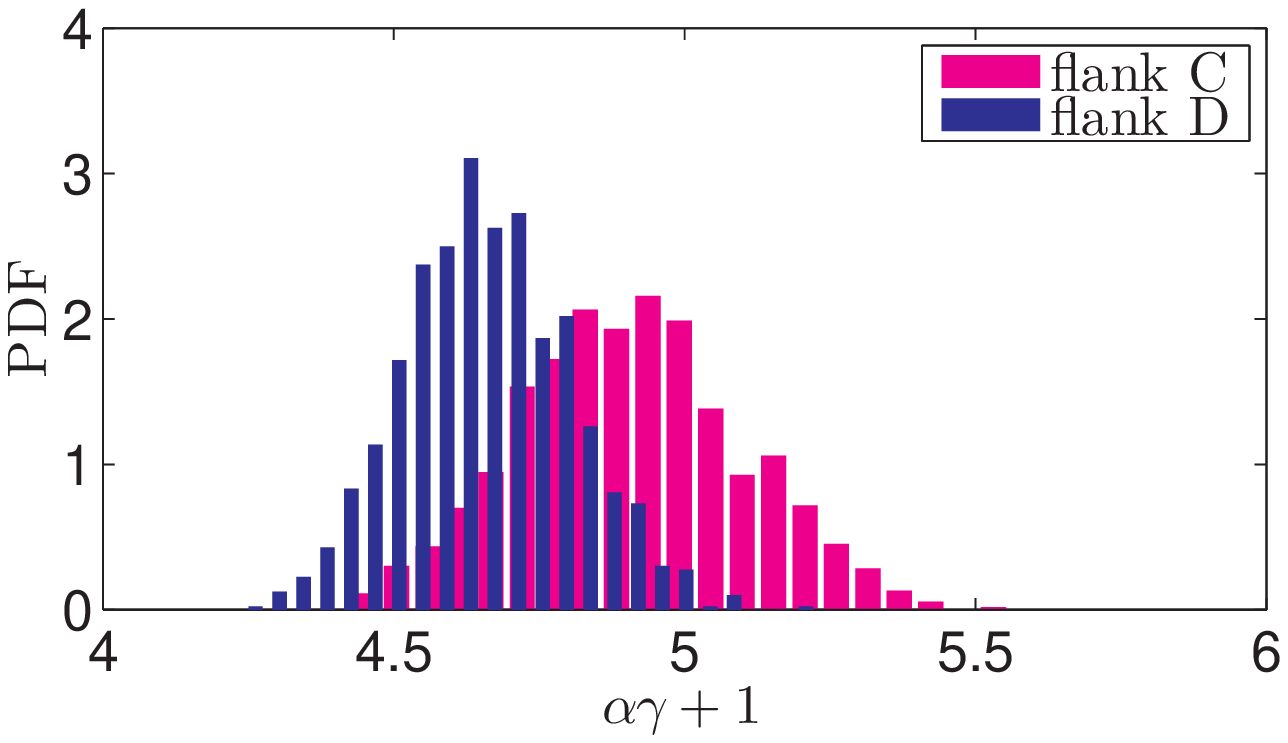} & 
\includegraphics[width = \myFigureWidth \textwidth]{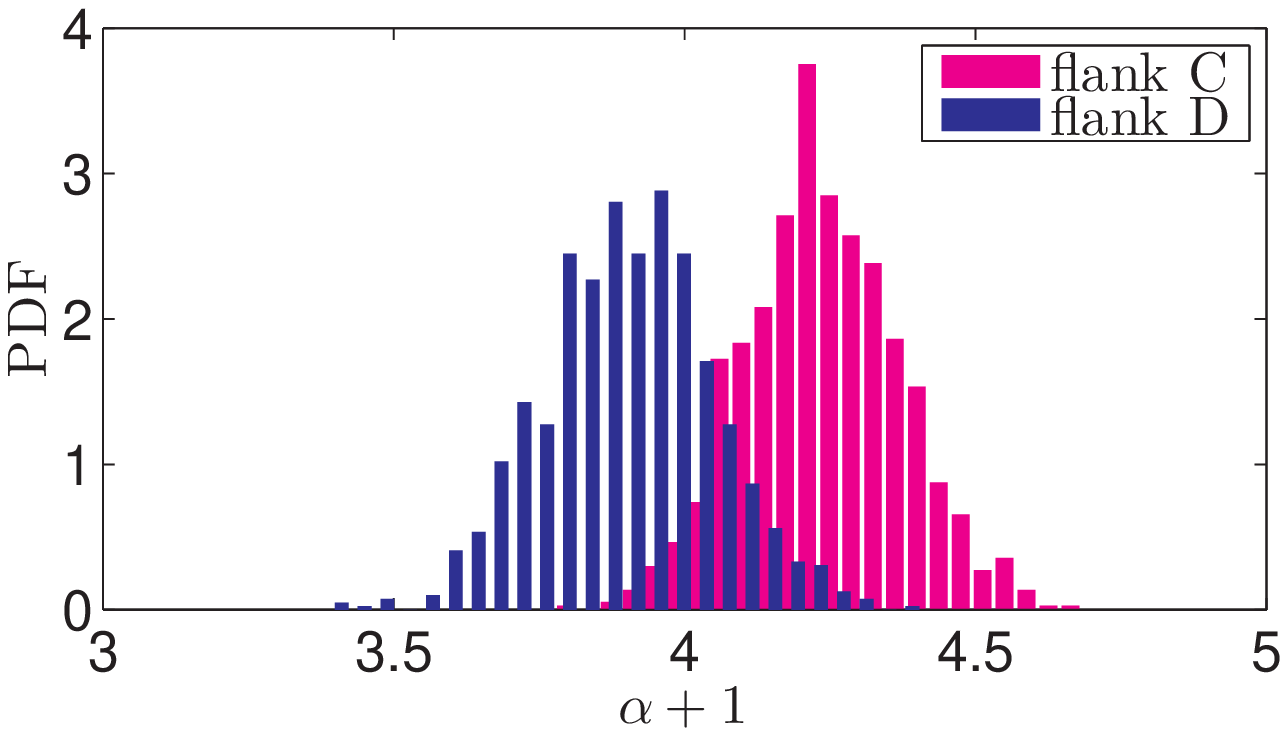} \\
\includegraphics[width = \myFigureWidth \textwidth]{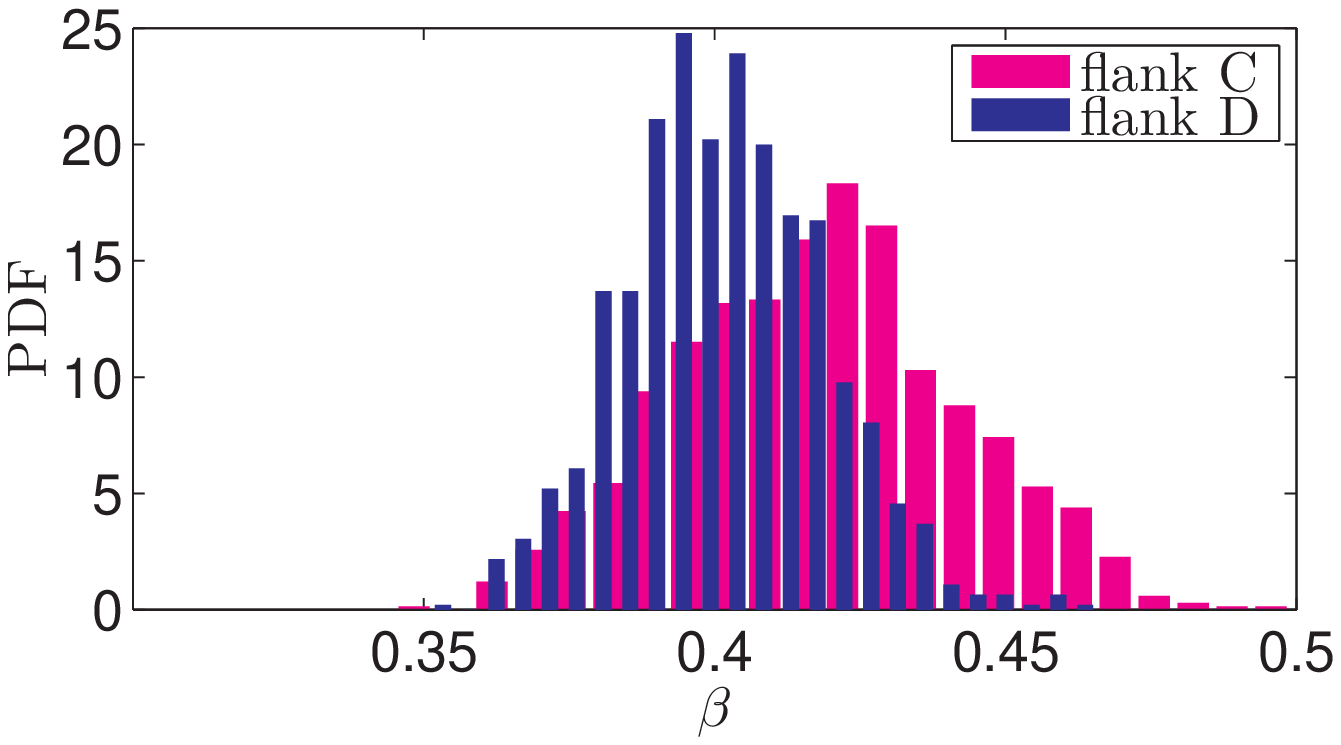} & 
\includegraphics[width = \myFigureWidth \textwidth]{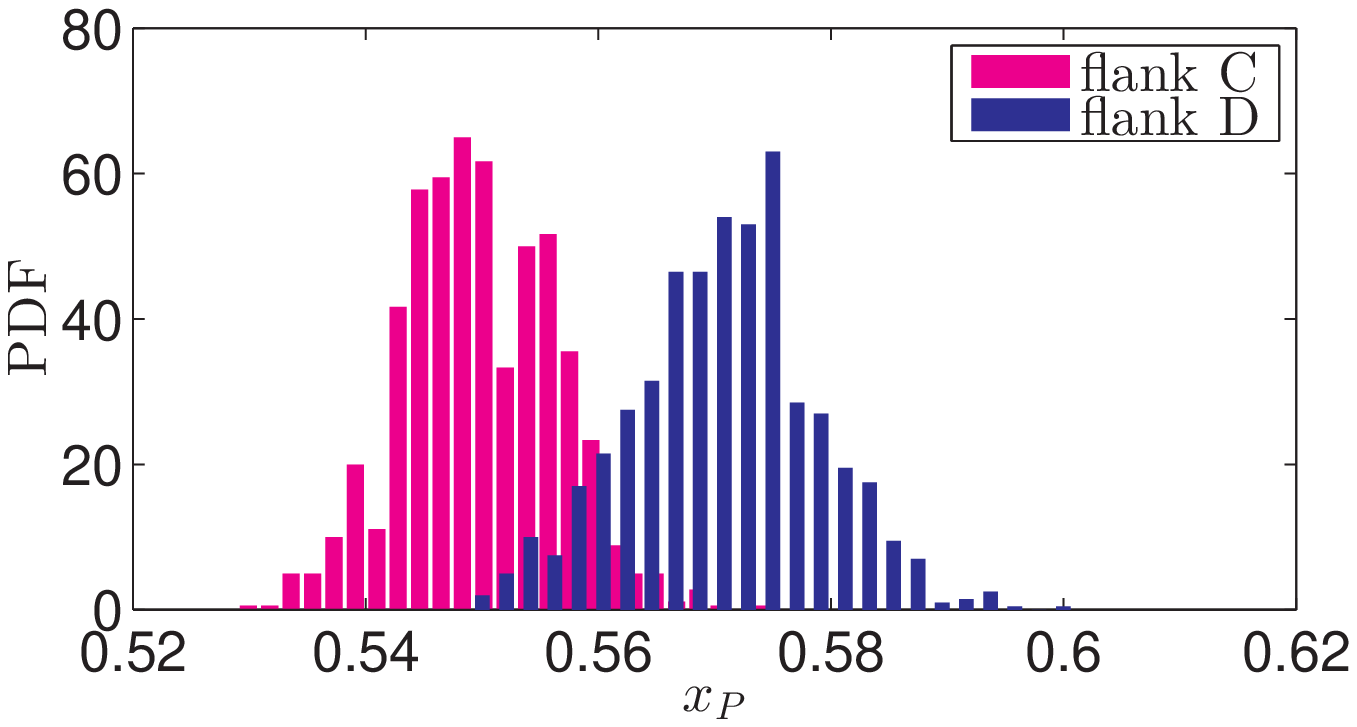} \\
\end{tabular}
\caption{Flanker test. Left column: GIGa; right column: LNP. Row one: fits of control
group; row two: fits of dyslexic group; row three: fits from first and second row
compared; row four: same as in row three, rescaled to unity mean; row five:
bootstrap-obtained distributions for power-law exponents ($\alpha\gamma+1$) and ($\alpha+1$);
row six: bootstrap-obtained distributions for scale parameters $\beta$ and $x_P$.}
\label{figure4}
\end{figure}

In Figure \ref{figure4}, the left column corresponds to GIGa fitting and the right column
to LNP fitting. From top to bottom, we present the distribution fits of the control and
dyslexic groups, followed by comparison of the two fitted distributions and then the
same two now rescaled to have a unity mean. We then show the bootstrap-obtained
distribution of $\alpha\gamma+1$ and $\beta$ for GIGa and of ($\alpha+1$) and $x_P$ for LNP. Clearly, the test
is quite consistent with rescaling and similarity of underlying neurophysiology.

\subsection{Arithmetic}

\begin{figure}[!htbp]
\centering
\begin{tabular}{cc}
\includegraphics[width = \myFigureWidth \textwidth]{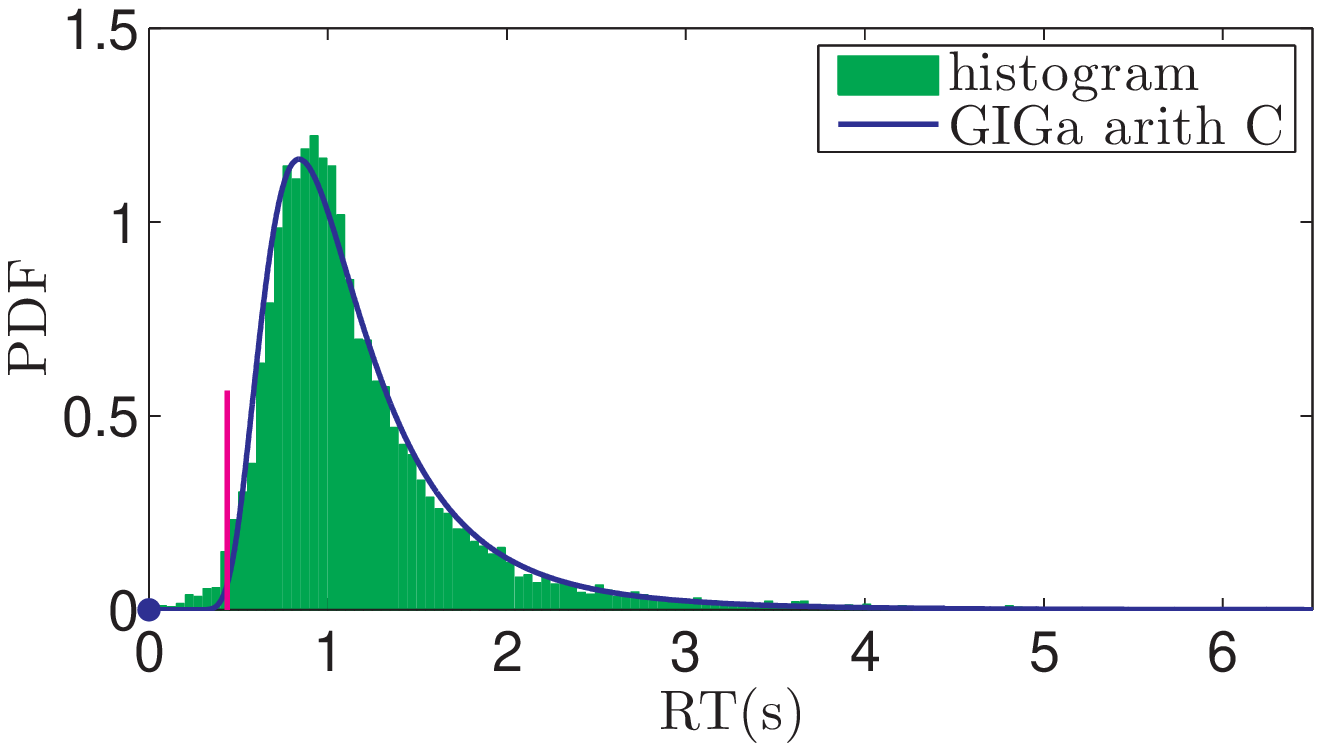} & 
\includegraphics[width = \myFigureWidth \textwidth]{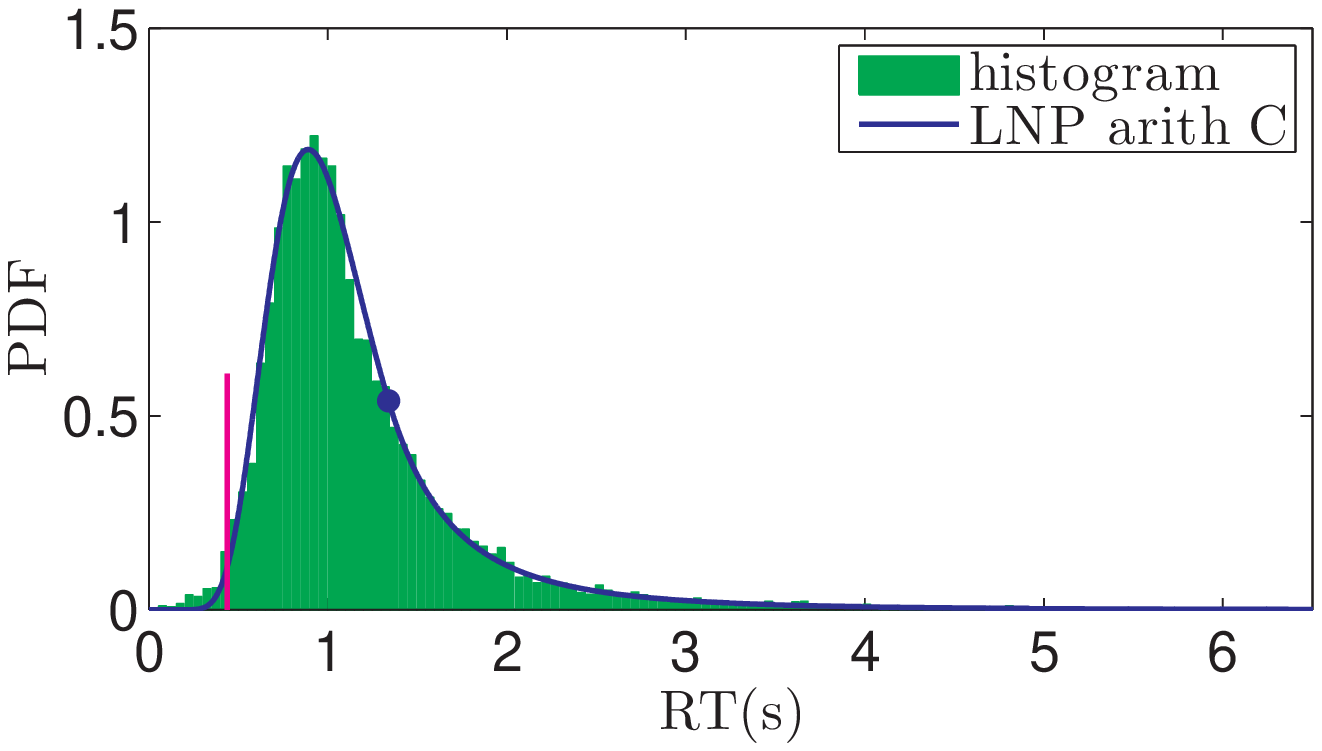} \\
\includegraphics[width = \myFigureWidth \textwidth]{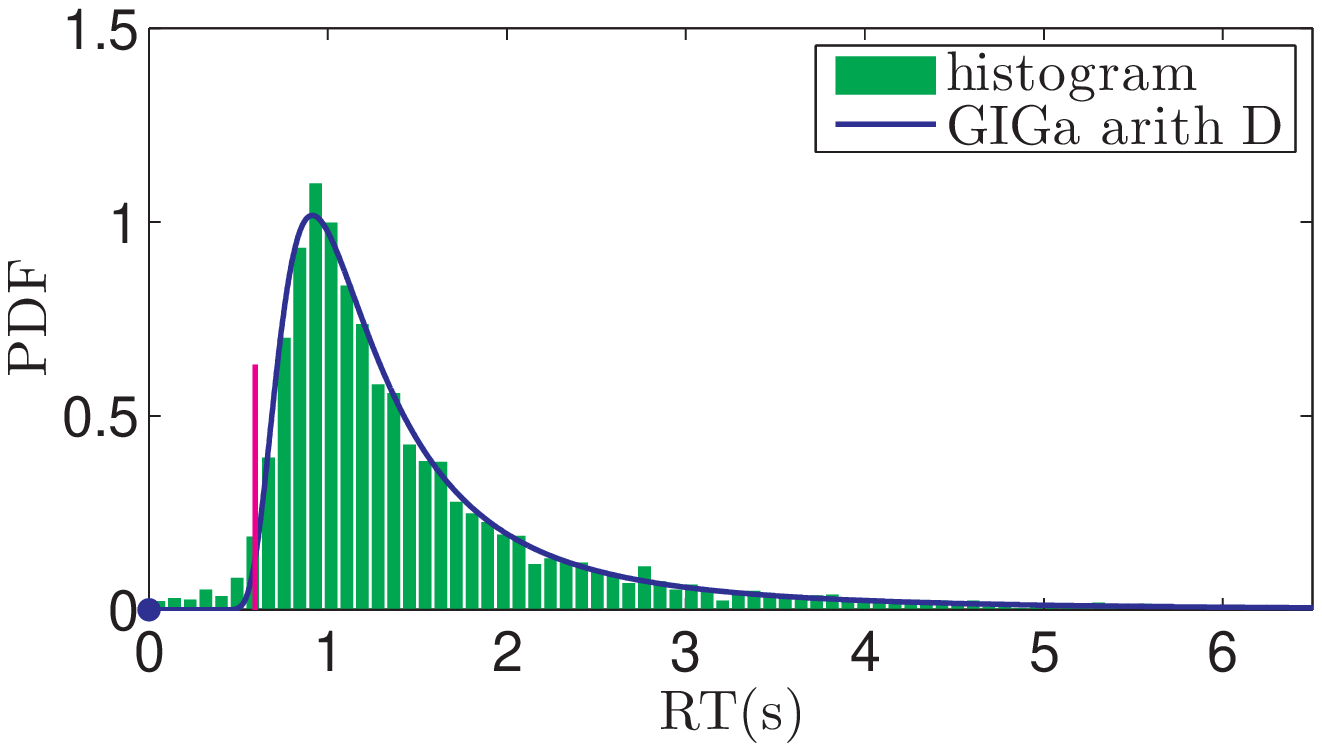} & 
\includegraphics[width = \myFigureWidth \textwidth]{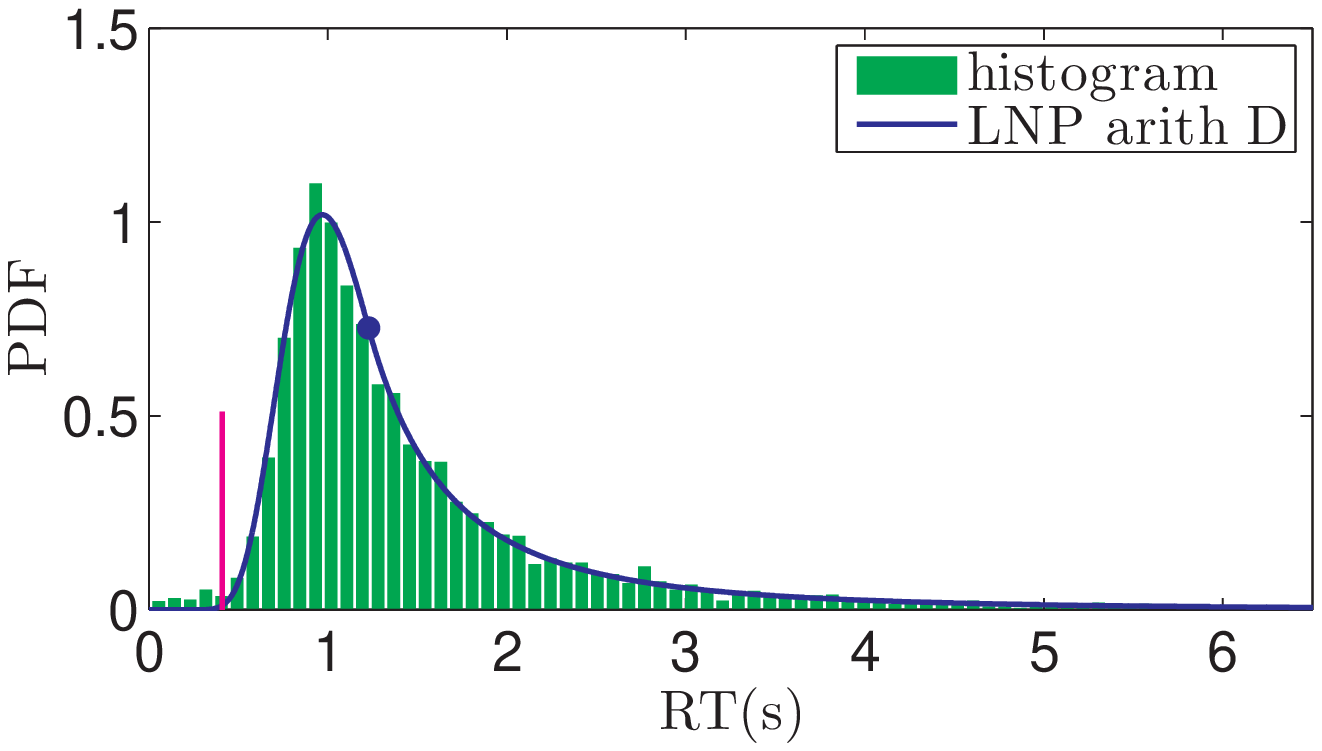} \\
\includegraphics[width = \myFigureWidth \textwidth]{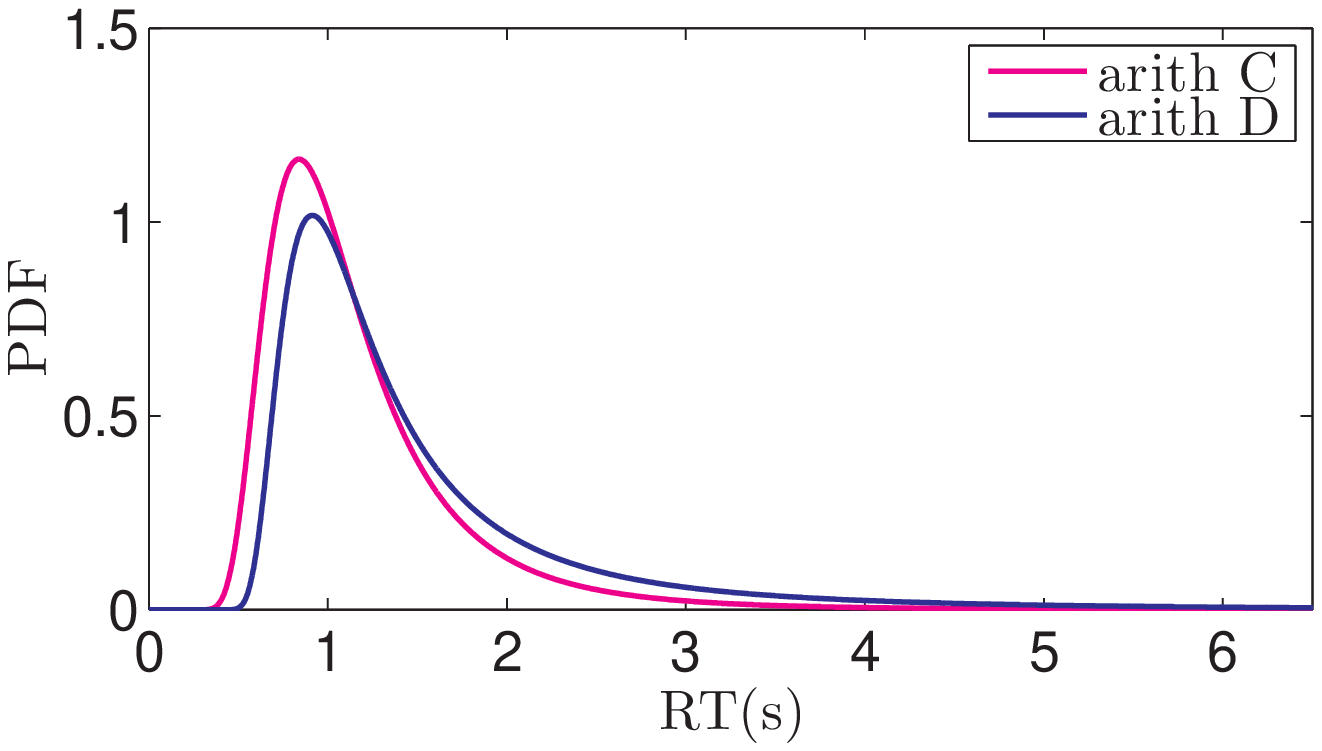} & 
\includegraphics[width = \myFigureWidth \textwidth]{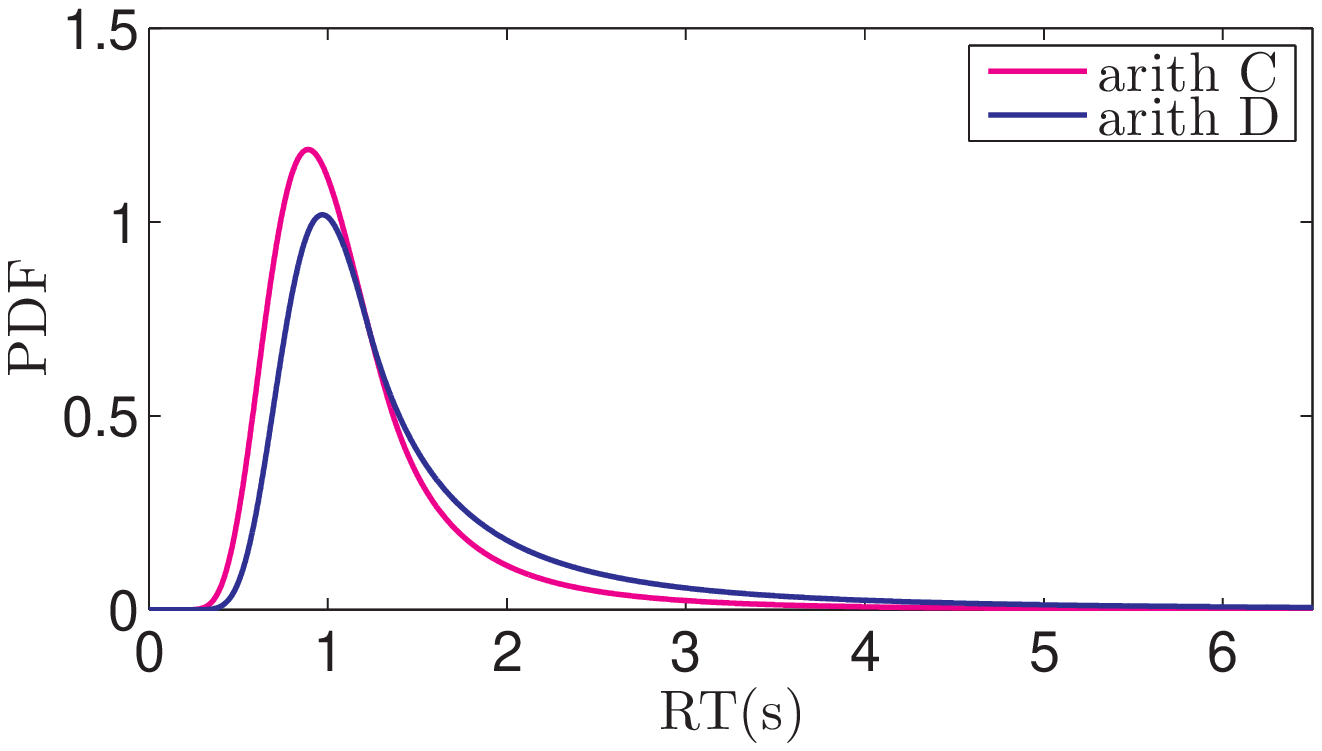} \\
\includegraphics[width = \myFigureWidth \textwidth]{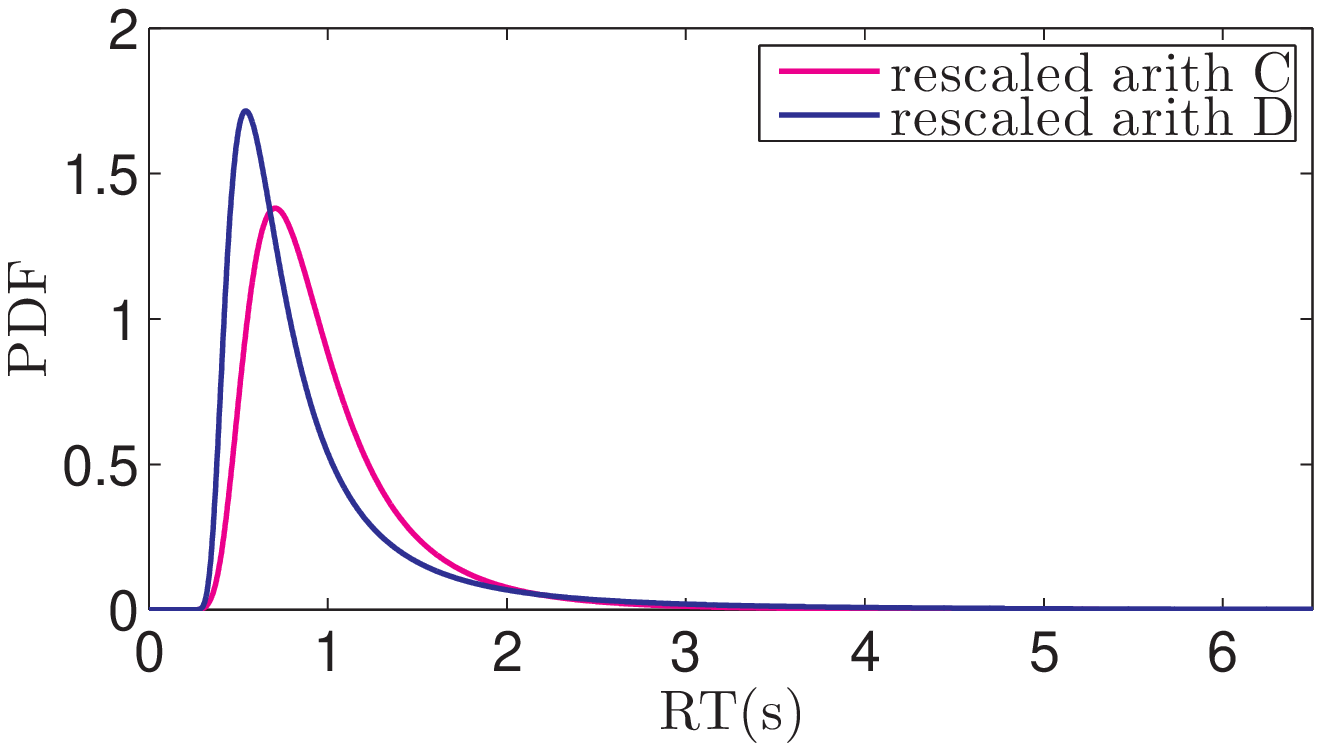} & 
\includegraphics[width = \myFigureWidth \textwidth]{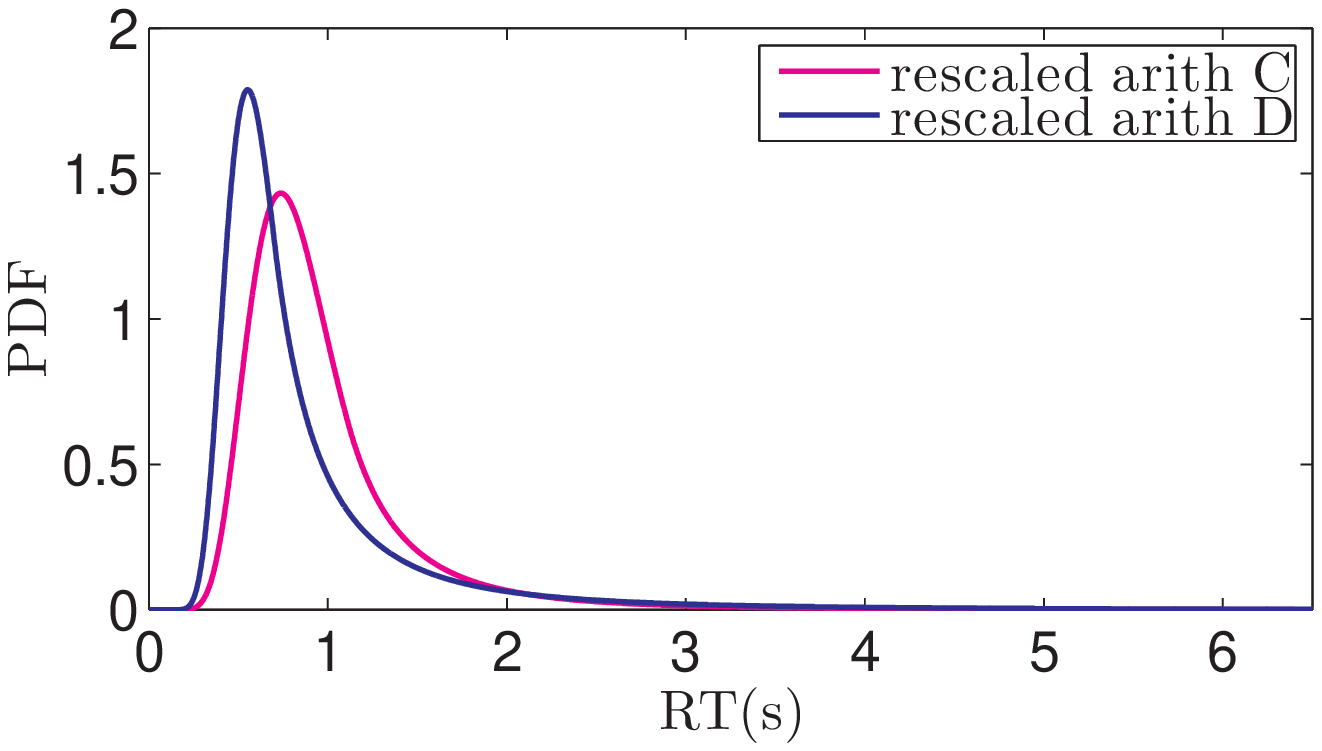} \\
\includegraphics[width = \myFigureWidth \textwidth]{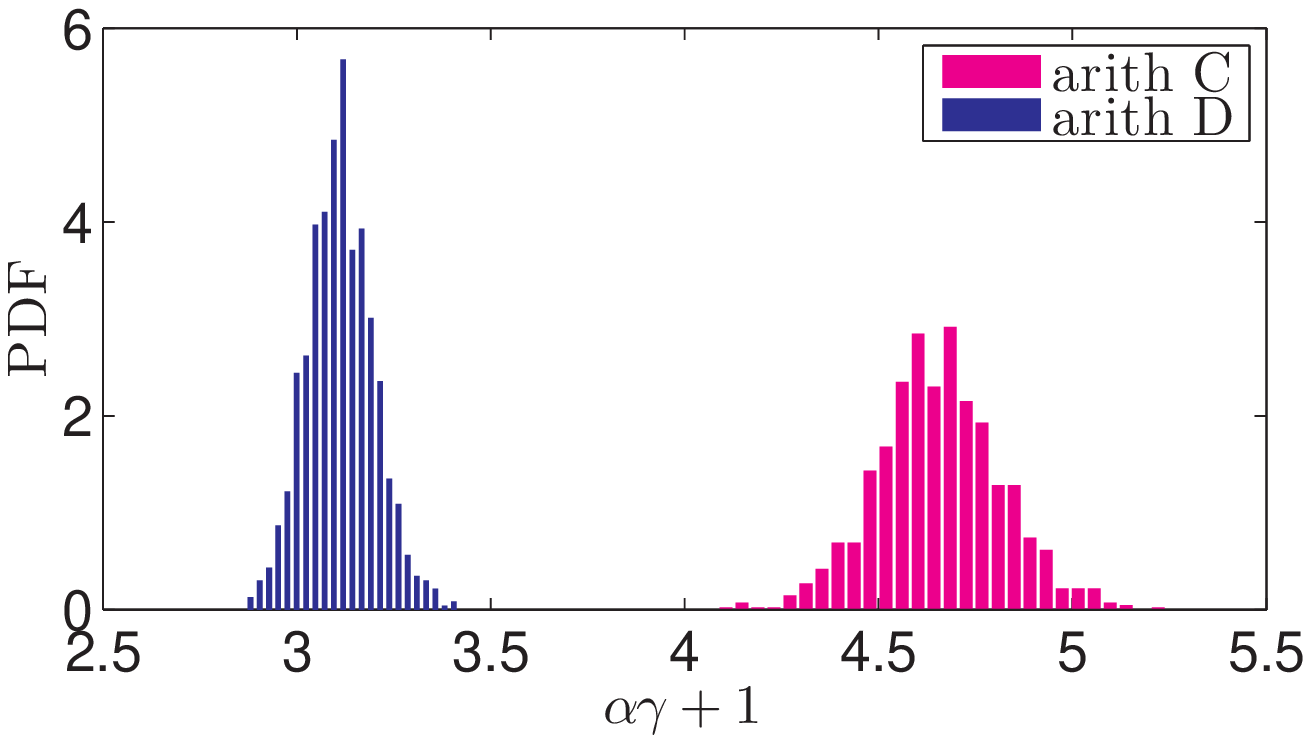} & 
\includegraphics[width = \myFigureWidth \textwidth]{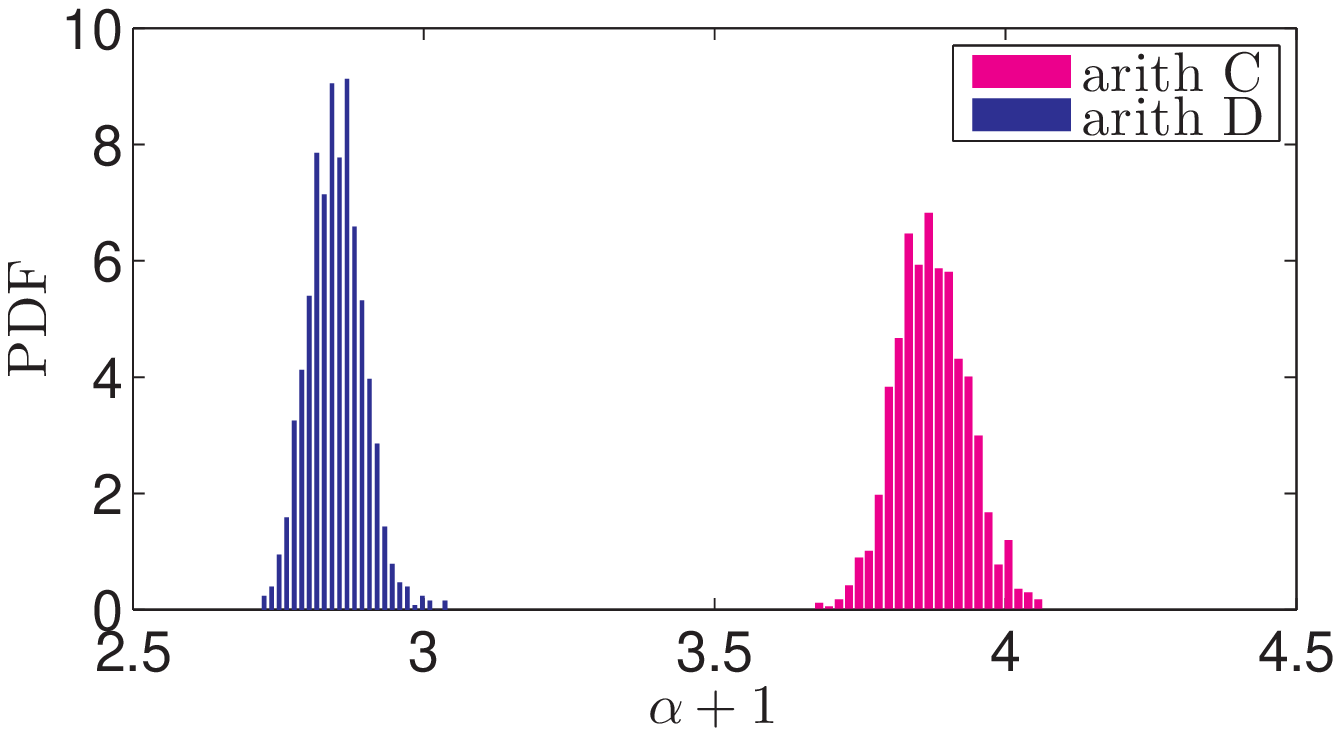} \\
\includegraphics[width = \myFigureWidth \textwidth]{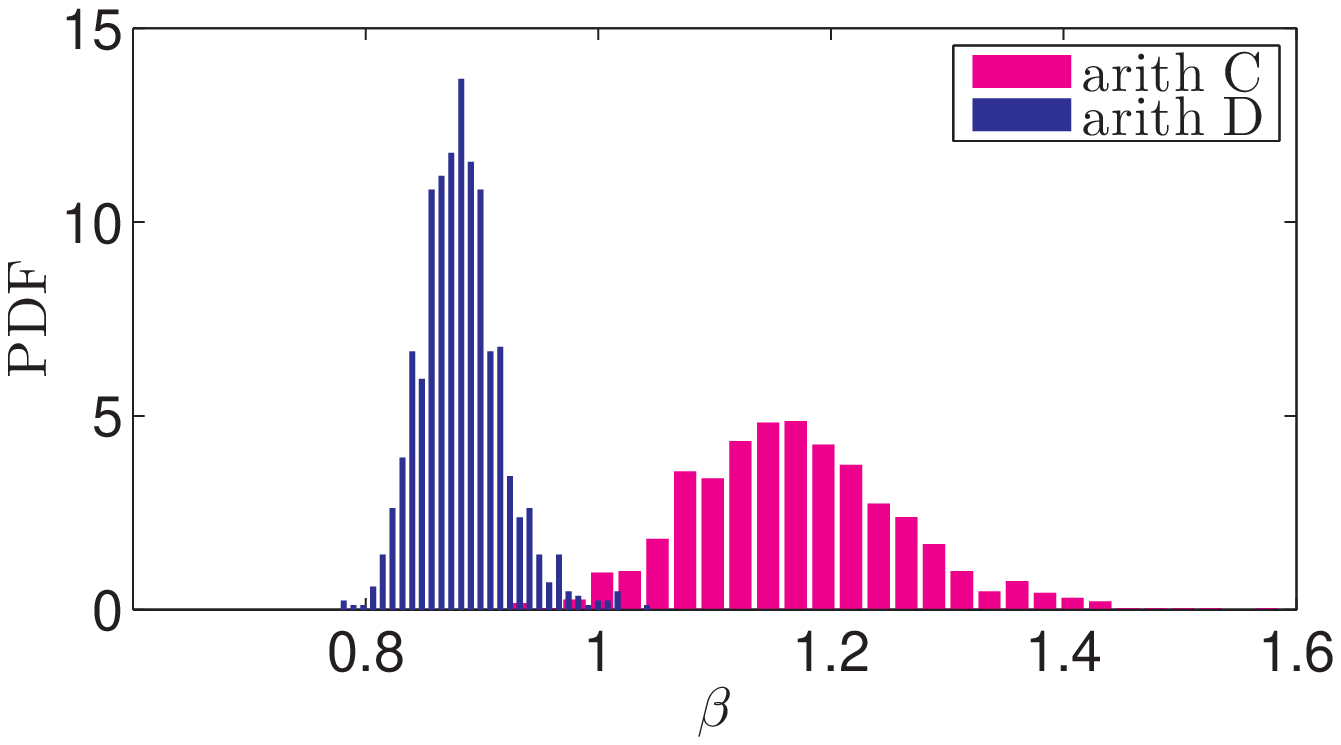} & 
\includegraphics[width = \myFigureWidth \textwidth]{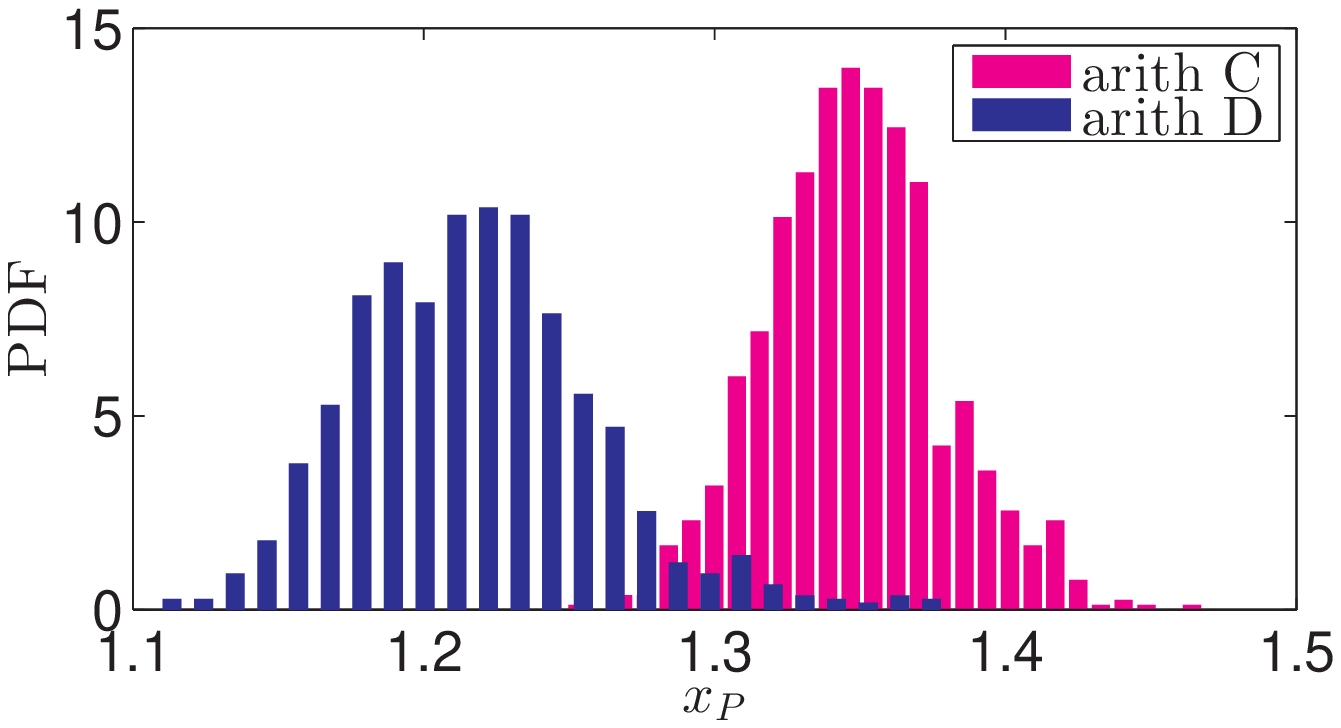} \\
\end{tabular}
\caption{Arithmetic test. Left column: GIGa; right column: LNP. Row one: fits of
control group; row two: fits of dyslexic group; row three: fits from first and
second row compared; row four: same as in row three, rescaled to unity mean; row
five: bootstrap-obtained distributions for power-law exponents ($\alpha\gamma+1$) and
($\alpha+1$); row six: bootstrap-obtained distributions for scale parameters $\beta$ and $x_P$.}
\label{figure5}
\end{figure}

We now present the same figure, Figure \ref{figure5}, for the arithmetic test. Additionally, in Table \ref{table5}, we show
confidence intervals for the power-law tail exponent and the scale parameters of
GIGa and LNP. 
For comparison purposes, we also give the confidence intervals for the flanker test in Table \ref{table6}.
Confidence intervals in Table \ref{table5} and \ref{table6} reflect the confidence level of $95\%$.

\begin{table}[!htbp]
\centering
\caption{Confidence intervals for arithmetic Test}
\label{table5}
\begin{tabular}{crl} 
\hline
\multicolumn{3}{c}{GIGA CI of ($\alpha\gamma+1$)} \\
\hline
Control  & (4.469, & 5.114) \\
Dyslexia & (2.948, & 3.291) \\
\hline
\end{tabular}
\begin{tabular}{crl} 
\hline
\multicolumn{3}{c}{LNP CI of ($\alpha+1$)} \\
\hline
Control  & (3.741, & 3.984) \\
Dyslexia & (2.783, & 2.994) \\
\hline
\end{tabular}
\begin{tabular}{crl} 
\hline
\multicolumn{3}{c}{GIGA CI of $\beta$} \\
\hline
Control  & (1.074, & 1.490) \\
Dyslexia & (0.817, & 0.949) \\
\hline
\end{tabular}
\begin{tabular}{crl} 
\hline
\multicolumn{3}{c}{LNP CI of $x_P$} \\
\hline
Control  & (1.277, & 1.391) \\
Dyslexia & (1.165, & 1.359) \\
\hline
\end{tabular}
\end{table}

\begin{table}[!htbp]
\centering
\caption{Confidence intervals for flanker Test}
\label{table6}
\begin{tabular}{crl} 
\hline
\multicolumn{3}{c}{GIGA CI of ($\alpha\gamma+1$)} \\
\hline
Control  & (4.530, & 5.281) \\
Dyslexia & (4.381, & 4.944) \\
\hline
\end{tabular}
\begin{tabular}{crl} 
\hline
\multicolumn{3}{c}{LNP CI of ($\alpha+1$)} \\
\hline
Control  & (3.984, & 4.512) \\
Dyslexia & (3.625, & 4.200) \\
\hline
\end{tabular}
\begin{tabular}{crl} 
\hline
\multicolumn{3}{c}{GIGA CI of $\beta$} \\
\hline
Control  & (0.370, & 0.465) \\
Dyslexia & (0.369, & 0.435) \\
\hline
\end{tabular}
\begin{tabular}{crl} 
\hline
\multicolumn{3}{c}{LNP CI of $x_P$} \\
\hline
Control  & (0.537, & 0.563) \\
Dyslexia & (0.554, & 0.587) \\
\hline
\end{tabular}
\end{table}

Notice that when the scale parameters "coincide" for two groups, the groups are still
different if the shape parameters are different. For instance, while the $x_P$ confidence
intervals overlap for the arithmetic test, those for ($\alpha+1$) do not (nor do they for 
HW, \citep[][in preparation]{liuInPreparation}).

Our analysis extends to the color and the word-naming tests, \citep{holden2014dyslexic} with
the former consistent with rescaling and the latter not \citep[][in preparation]{liuInPreparation}.
We should point out that, unlike word-naming, the arithmetic result is entirely
unexpected. 

\section{Conclusions}

We discussed properties of the candidate distributions with "fat" power-law tails:
stable, generalized inverse gamma and mixture lognormal-Pareto. We emphasized
their shape and scale parameters, as those are crucial for interpreting similarities
and differences in the underlying neurophysiology of cognition between a group of
interest and a control group.

We hypothesized that when the response time distributions of the two groups can
be rescaled to each other by the ratio of the mean responses, the supporting
cognitive and neurophysiological organization should be interpreted as generally
equivalent, but proportionally stretched or compressed in the temporal domain. For
the candidate distributions (and other analytical distributions), such rescaling is
achieved by rescaling of the scale parameter.

Conversely, absent the rescaling between the distributions, the neurophysiology
should be generally interpreted as different. For the candidate distributions (and
other analytical distributions), the difference is expressed via the difference of the
shape, or shape-related, parameters.

In order to improve statistics for analysis of the similarity and difference between
the two groups, we combined all the subjects of each group into a single set. This
deviates from the common practice in which each subject is fitted separately and the
parameters of the fits are subsequently averaged. We conjectured that each subject 
may be interpreted as a random variate of the response time distribution, the latter 
corresponding to a particular neurophysiological process. We conducted a numerical 
simulation to successfully test our conjecture.


We illustrated our approach on the response-time trials with dyslexic children and
their control counterparts. We showed that, using the combined sets of each group
for the analysis, the distributions of the dyslexic and control groups are consistent
with rescaling for simpler cognitive tasks but not so for higher-level tasks. We used
the distribution's half-width and power-law tail exponent as shape parameters and
bootstrap to obtain confidence intervals. We hope to extend our approach to clinical
trials, such as studies of efficacy of ADHD medications \citep{epstein2011effects}.

\section*{References}

\bibliography{mybibfile}

\end{document}